\documentclass[aps,prx,twocolumn,longbibliography,superscriptaddress,floatfix,nofootinbib]{revtex4-2}
\usepackage[utf8]{inputenc}
\usepackage[english]{babel}
\usepackage{amsfonts}
\usepackage{lipsum}
\usepackage{graphicx}
\usepackage{float}
\usepackage{bm}
\usepackage{xcolor}
\usepackage{tcolorbox}
\usepackage{epstopdf}
\usepackage{amsmath}
\usepackage{amssymb}
\usepackage{epstopdf}
\usepackage[normalem]{ulem} 
\usepackage{enumerate}
\usepackage{tikz}
\usepackage[urlcolor=blue,colorlinks=true,citecolor=red,linkcolor=red,pdfstartview={FitH},bookmarks=false]{hyperref}
\usepackage{lipsum}

\newcommand\doublecheck{\textcolor{black}{\checked\kern-0.6em\checked}}

\newcommand{\bs}{\boldsymbol}
\newcommand{\nn}{\nonumber}

\graphicspath{{fig/}{./fig/}{.}}

\DeclareMathOperator{\sgn}{sgn}
\makeindex

\begin{document}
\newcommand{\LUND}{\affiliation{Solid State Physics and NanoLund, Lund University, Box 118, S-221 00 Lund, Sweden}}
 \newcommand{\WARSAW}{\affiliation{International Research Centre MagTop, Institute of Physics, Polish Academy of Sciences, Aleja Lotnikow 32/46, PL-02668 Warsaw, Poland}}
 \newcommand{\HAMBURG}{\affiliation{The Hamburg Centre for Ultrafast Imaging, Luruper Chaussee 149, 22761 Hamburg, Germany}}
\newcommand{\GERMANY}{\affiliation{I. Institut für Theoretische Physik, Universität Hamburg, Notkestra{\ss}e 9, 22607 Hamburg, Germany}}
\newcommand{\DELHI}{\affiliation{Department of Physics and Astrophysics, University of Delhi, Delhi-110007, India}}

\newcommand{\COFIRST}{\thanks{OAA, II and AM contributed equally to this work.}}

\title{Controlling Majorana hybridization in magnetic chain-superconductor systems}

 \author{Oladunjoye A. Awoga}
\COFIRST
\LUND
\email[e-mail:]{oladunjoye.awoga@ftf.lth.se}

\author{Ioannis Ioannidis} 
\COFIRST
\HAMBURG
\email[e-mail:]{ioannis.ioannidis@uni-hamburg.de}
\GERMANY

\author{Archana Mishra} 
\COFIRST
\WARSAW
\DELHI
\email[e-mail:]{amishra@physics.du.ac.in}

\author{Martin Leijnse}
\LUND

\author{Mircea Trif}
\WARSAW
	
\author{Thore Posske}
\HAMBURG
\GERMANY

\date{\today} 

\begin{abstract}
We propose controlling the hybridization between Majorana zero modes at the ends of magnetic adatom chains on superconductors by an additional magnetic adatom deposited close by.
By tuning the additional adatom's magnetization, position, and coupling to the superconductor, we can couple and decouple the Majorana modes as well as control the ground-state parity.
The scheme is independent of microscopic details in ferromagnetic and helical magnetic chains on superconductors with and without spin-orbit coupling, which we show by studying their full microscopic models and their common low-energy description.
Our results show that scanning tunneling microscopy and electron spin resonance techniques are promising tools for controlling the Majorana hybridization in magnetic adatoms-superconductor setups, providing a basis for Majorana parity measurements, fusion, and braiding techniques.
\end{abstract}
	
\maketitle

 %
\section{Introduction}
 Majorana zero modes (MZMs) are non-Abelian quasiparticles that are their own anti-particles and have been subject to intense research in condensed matter physics due to their exotic properties and potential application in fault-tolerant topological quantum computing~\cite{kitaev2003fault,nayak2008non,Leijnse2012Introduction,pachos2012introduction,alicea2012new,stern2013topological,Posske2020,Marra2022Majorana}. Among the most promising platforms for realizing MZMs are systems based on topological insulators ~\cite{fu2008superconducting,fu2009josephson,li2014majorana}, fractional quantum Hall systems~\cite{moore1991nonabelions,sarma2005topologically,willett2013magnetic}, ultracold atoms~\cite{shan2014majorana,qu2015majorana,ptok2018quantum}, semiconducting nanowires~\cite{lutchyn2010majorana,oreg2010helical,mourik2012signatures,beenakker2013search}, planar Josephson junctions \cite{Hell2017Two,Pientka2017Topological,Ren2019Topological,Fornieri2019Evidence} and magnetic adatoms chain deposited on a superconductor~\cite{nadj2013proposal,klinovaja2013topological,Pientka2013topological,Pientka2014Unconventional,nadj2014observation,brydon2014topological,li2014topological,andolina2017topological,howon2018Toward,schneider2021topological,Kobialka2021Majorana}. Although there has been immense theoretical and experimental progress, especially in semiconducting nanowires-superconducting hybrid structures, the short nanowire length and disorder in these systems are a major challenge in the search for MZMs \cite{das2023search}. Progress with this platform demands clean systems with nanowires much longer than the coherence length of the superconductor, which is difficult with present technology.
 In the search for alternatives, magnetic adatoms on a superconductor have garnered a lot of interest, where MZMs could be directly probed using a scanning tunneling microscope~\cite{nadj2014observation,ruby2015end,Jeon2017Distinguishing,jack2021detecting}.
Here, a single magnetic adatom deposited on a superconductor induces a spin-polarized Yu-Shiba-Rusinov (YSR) state within the superconducting gap~\cite{Yu1965Bound,shiba1968classical,rusinov1969theory,menard2015coherent}. When a chain of such magnetic adatoms is deposited on a superconductor, the spin-polarized YSR states from the individual magnetic adatoms overlap to form a nondegenerate Shiba band, which leads to an effective $p$-wave superconducting gap in the chain's electronic spectrum when an effective spin-orbit coupling is present. This $p$-wave superconductor formed by the magnetic adatom chain is topologically nontrivial, hosting a MZM at each end.  Two salient frameworks to realize this proposal are a chain of magnetic adatoms with helical magnetization on an $s$-wave superconductor~\cite{nadj2013proposal,klinovaja2013topological,Pientka2013topological,Pientka2014Unconventional,choy2011majorana,nakosai2013two,braunecker2013interplay,vazifeh2013self,poyhonen2014majorana,reis2014self,weststrom2015topological,peng2015strong,hoffman2016topological,li2016manipulating,neupert2016shiba} and a chain of ferromagnetic adatoms deposited on an $s$-wave superconductor with spin-orbit coupling~\cite{li2014topological,nadj2014observation,brydon2014topological,ruby2015end,rontynen2015topological,andolina2017topological,howon2018Toward,schneider2021topological,Schneider2022}. 

Several STM experiments have reported zero bias peaks in differential conductance measurements on magnetic chain-superconductor systems, which have been interpreted in terms of MZMs~ \cite{nadj2014observation,ruby2015end,howon2018Toward,schneider2021topological}. However, zero-bias peaks can also stem from trivial in-gap states in the magnetic chain~\cite{Hess2022Prevalence,kuster2022non}. In fact, it has been established across realistic platforms proposed for creating MZMs that the observed zero-bias conductance peaks can have alternative explanations~\cite{Awoga2019Supercurrent,Prada2020From,Zhang2020Transport,Hess2022Prevalence}, a development that has opened up a debate about whether the MZMs have been observed in reported experiments.
Unambiguous verification of zero-bias peaks as MZM signals involves measurements exploring their non-Abelian properties, such as fusion and braiding behavior, which are not shared by trivial in-gap states, or via cross-correlation shot noise measurements~\cite{manousakis2020weak}. Although there is an abundance of such proposals in semiconducting platforms, see Refs.~\cite{vanHeck2012Coulomb,hyart2013flux,Amorin2015Majorana,aasen2016milestones,hell2016time,clarke2017probability,zhou2022fusion,seoane2022fusion,liu2022fusion,tsintzis2023roadmap} for examples, the adatom route has received much less attention ~\cite{li2016manipulating,bedow2023implementation}. This can be partly attributed to the rigidity of this platform since, once fabricated, the system parameters are difficult to manipulate externally, which is a prerequisite for quantum information processing. To carry out non-Abelian experiments, it is imperative to have a high-quality, time-dependent control over the hybridization between MZMs, resulting in their coupling and decoupling, as a first step towards experiments revealing their non-Abelian character.

In this paper, we propose a generic method to control and shield, i.e., switch off, the hybridization of MZMs in magnetic chain-superconductor systems via a single additional magnetic adatom. We consider two chains of magnetic adatoms on an $s$-wave superconductor with a single magnetic adatom in between such that there is finite overlap between the MZMs at the ends of the magnetic chains in the topological phase and the YSR state induced by the single magnetic adatom. We find a model-independent strong tunability of the MZMs hybridization energy by changing the single adatom's magnetic orientation, its distance from the magnetic chains and its coupling to the superconductor. A tunability of the Majorana hybridization by magnetism at the quantum spin Hall edge has been described theoretically in Ref.~\cite{Keidel2018Tunable}.
Here, the control of the parameters of the single magnetic adatom can, in principle, be achieved using electron spin resonance scanning tunneling microscopy (ESR-STM) techniques~\cite{Balatsky2012Electron,Baumann2015Electron,Yang2019Tuning,Drost2021Combining}, see Sec.~\ref{sec:Expt} for details.
Remarkably, we find that these parameters tune the MZMs hybridization through a vast range of values: between strongly hybridized MZMs and zero, where the MZMs are perfectly shielded from each other.
The complete shielding of the MZMs occurs for fine-tuned parameters that mark a phase transition between regions with different ground-state parity. Here, ``fine-tuned" refers to the points where the decoupling is exact. However, approximately decoupled MZMs suffice for performing non-Abelian operations on the appropriate time-scales, see Section~\ref{sec:Expt} and Eq.~\eqref{eq:OpTime} for details.
%
We show that our results are general and not model specific, which we demonstrate by the general low-energy model and specific well-studied microscopic models of helical and ferromagnetic adatoms chains in both continuum and lattice superconductor models. As an overarching theme, our results show a wide variability of the MZMs hybridization with the orientation of the single magnetic adatom, which is accessible with current experimental techniques.
%
Control of MZM hybridization is an essential prerequisite for Majorana fusion and braiding experiments. Thus, our work is a step towards realizing these MZMs prospects and opens a novel experimentally realizable research direction in the magnetic chain-superconductor platform.
Moreover, the shielding effect can be used to decouple strongly hybridized Majorana modes, also called precursors of Majorana modes, which were observed in short magnetic chains~\cite{Schneider2022}.
Furthermore, changes in the ground state parity imply control over the occupation of the ground state of the system, which is essential for initialization in parity-based Majorana correlation measurements, where perturbing one of the chains induces measurable parity oscillation in the other chain~\cite{Burnell2013Measuring,Burnell2014Correlated,San-Jose2012ac}.
Our results enhance the understanding of the behavior of MZMs and inspire future experiments for probing non-Abelian properties of MZMs and employing them in fault-tolerant quantum computing.

The remainder of the paper is organized as follows: In Sec.~\ref{sec:Sys}, we introduce the system to control the MZMs hybridization and present a general low-energy model to analytically evaluate the effective Majorana hybridization. We show the possible decoupling of the MZMs due to the overlap with the YSR state from the single magnetic adatom and also provide an analysis of the change in the ground state parity. In Sec.~\ref{sec:Control}, we show the generality of the control of MZMs hybridization by addressing three experimentally relevant physical realizations. We first present general analytical results for continuum superconductors in Sec.~\ref{subsec:Cont}, and then consider helically ordered magnetic chains on a conventional $s$-wave superconductor and ferromagnetic ordered chains on an $s$-wave superconductor with non-zero spin-orbit coupling both in the continuum limit, in Sec.~\ref{subsubsec:PientkaModel} and Sec.~\ref{subsubsec:BrydonModel}, respectively. In Sec.~\ref{sec:LatticeModel}, we consider a ferromagnetic chain on a square lattice superconductor with finite spin-orbit coupling.
In Sec.~\ref{sec:Expt}, we discuss a possible experimental protocol for the proposed setup and the adiabatic constraints.
Finally, we give concluding remarks in Sec.~\ref{sec:Conc}, where we discuss possible future experiments achievable with our proposal, including manipulating and decoupling precursors of Majorana modes, and a possible method to control the Majorana hybridization using the dynamics of the single magnetic adatom. 
\section{The System}\label{sec:Sys}
\begin{figure*}[!t]
   \centering
\includegraphics[width=.66\textwidth,trim= {5cm 18.5cm 15cm 2.5cm},clip]{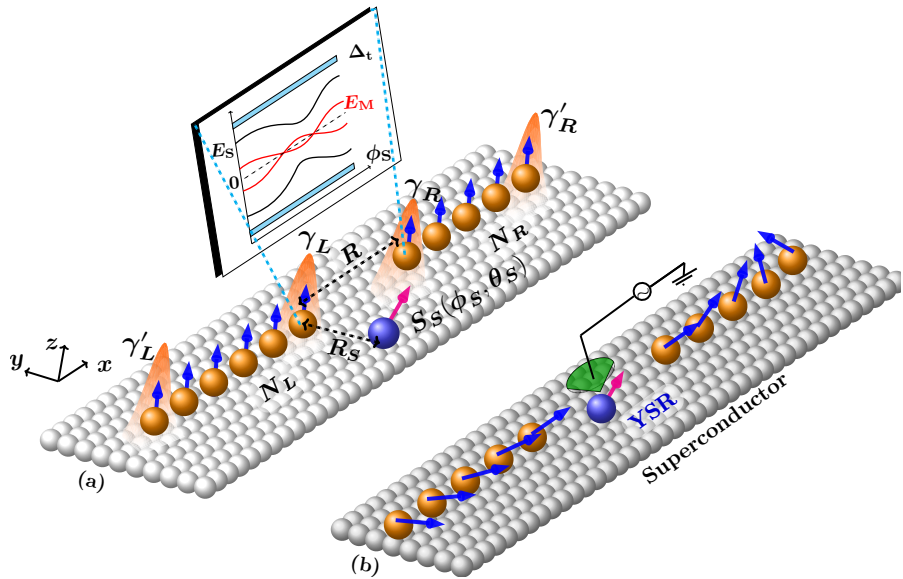}
\caption{Realizations of the system. (a) Two ferromagnetic chains with the left and right chain consisting of $N_{\rm L}$ and  $N_{\rm R}$ adatoms, respectively, on an $s$-wave superconductor, separated by a distance $R$. An additional adatom (blue) with magnetic orientation $\bs S_{\rm S}^{}(\phi_{\rm S}^{},~\theta_{\rm S}^{})$ is placed at distance $\bs R_{\rm S} =\left(x_{\rm S},\,y_{\rm S},\,z_{\rm S}\right)$ from the left chain. In the topologically nontrivial phase, the MZMs $\gamma_{\rm L}^\prime$, $\gamma_{\rm L}^{}$, $\gamma_{\rm R}^{}$, and $\gamma_{\rm R}^\prime$ reside at the chains' ends, whose density of states is shown in orange. Inset: schematic dependence of the effective MZMs hybridization energy $E_{\rm M}$ and the effective YSR energy $E_{\rm S}^{}$ on $\phi_{\rm S}^{}$, $\Delta_{\rm t}$ denotes the topological gap. (b) Same as (a) but with helically magnetized  adatoms on an $s$-wave superconductor with vanishing spin-orbit coupling. The STM tip (green) can be used to control the parameters of the single magnetic adatom by applying a time-dependent electric field.}
\label{fig:Fig1}
\end{figure*}

We consider a setup with two chains, left and right, consisting of $N_{\rm L}$ and $N_{\rm R}$ magnetic adatoms, respectively, that are deposited on an $s$-wave superconductor and separated by a distance $R$ from each other, see Fig.~\ref{fig:Fig1}, where Fig.~\ref{fig:Fig1}(a) and (b) depict ferromagnetic chains and helical magnetic chain on an s-wave superconductor with finite and vanishing Rashba spin-orbit coupling, respectively.
In addition, a single magnetic adatom (blue sphere), with magnetic orientation
$\bs S_{\rm S} =  \bs S_{\rm S} (\sin\theta_{\rm S}^{}\cos\phi_{\rm S}^{},\sin\theta_{\rm S}^{}\sin\phi_{\rm S}^{},\cos\theta_{\rm S}^{})$,
where
$\theta_{\rm S}^{}$ and $\phi_{\rm S}^{}$
are the polar and azimuthal angle, respectively,
is placed at position
$ \bs R_{\rm S}^{}=(x_{\rm S}^{},y_{\rm S}^{},z_{\rm S}^{})$ relative to the right end of the left chain, inducing a YSR state of energy $\epsilon_{\rm S}^{}(\phi_{\rm S}^{},~\theta_{\rm S}^{})$ centered at this position.
The chains are in the topologically non-trivial phase, each accommodating a pair of MZMs at its ends which are represented by $\gamma$-operators, orange density of states in Fig.~\ref{fig:Fig1}. The magnetic chains are long enough to avoid overlap of the inner MZMs, $\gamma_{\rm L}$ and $\gamma_{\rm R}^{}$,  with the outer MZMs,
$\gamma_{\rm L}^\prime$ and $\gamma_{\rm R}^\prime$.
Yet, there is a finite bare hybridization energy,
$\epsilon_{\rm M}^{}\left( R\right)$,
between the inner MZMs,
$\gamma_{\rm L}^{}$ and $\gamma_{\rm R}^{}$,
leading to the formation of a non-local fermion with energy
$\epsilon_{\rm M}^{}$.
%
%
The superconductor mediates a finite coupling between the inner MZMs and the YSR state at $\bs R_{\rm S}^{}$ which effectively modifies both
$\epsilon_{\rm M}\left(R\right)$ and 
$\epsilon_{\rm S}^{}\left(\theta_{\rm S}^{},\phi_{\rm S}^{}\right)$ to new values
$E_{\rm M}^{}\left(R_{\rm S}^{},\theta_{\rm S}^{},\phi_{\rm S}^{},\epsilon_{\rm S}^{},\epsilon_{\rm M}^{},R\right)$
and
$E_{\rm S}^{}\left(R_{\rm S}^{},\theta_{\rm S}^{},\phi_{\rm S}^{},\epsilon_{\rm S}^{},\epsilon_{\rm M}^{},R\right)$,
respectively. 
A typical dependence of  $E_{\rm M}$ and $E_{\rm S}^{}$ on $\phi_{\rm S}^{}$ is shown in the inset of  Fig.~\ref{fig:Fig1}(a), explained in detail in Sec.~\ref{sec:Control}. It should be noted that $E_{\rm M}^{}$ and $E_{\rm S}^{}$ correspond to hybridized Majorana and YSR states with density of states at both original states' positions.

%
%
Assuming that $|\epsilon_{\rm M}|$ and $|\epsilon_{\rm S}|$ are much smaller than the topological gap of the Shiba bands, $\Delta_{\rm t}$, all systems discussed in this work share the same low-energy limit, where only the in-gap states in the central region of the setup are relevant, including the overlapping inner MZMs $\gamma_L$ and $\gamma_R$, and the single magnetic adatom's YSR state, see Fig.~\ref{fig:Fig1}.  We ignore the outer MZMs because they decouple from the central region for sufficiently long chains. After integrating out the superconductor and the Shiba bands of the chains, the effective low-energy Hamiltonian for the $\gamma_L$-YSR-$\gamma_R$ system is
\begin{equation}\label{eq:EffectiveH}
H_{\rm eff}=\frac{i}{2} \epsilon_{\rm M}^{} \gamma_{\rm L}^{} \gamma_{\rm R}^{}+\epsilon_{\rm S}^{} d^\dagger d+ \left[t_{\rm L}^{} d^\dagger \gamma_{\rm L}^{}+t_{\rm R}^{} d^\dagger \gamma_{\rm R}^{}+ \rm{h.c.}\right],
\end{equation}
 where $d^\dagger$ creates an electron in the YSR state,
 $t_{\rm L/R}=t_{\rm L/R}\left(R_{\rm S}^{},\theta_{\rm S}^{},\phi_{\rm S}^{},\epsilon_{\rm S}^{},\epsilon_{\rm M}^{},R\right)$
 is the coupling amplitude between the YSR state and the MZM $\gamma_{\rm L/R}$ with 
 $\{\gamma_i,\gamma_j\}=2\delta_{ij}$. 
 The Hamiltonian in Eq.~\eqref{eq:EffectiveH} preserves particle-hole symmetry,  breaks time-reversal symmetry, and thus belongs to the topological class $D$ in the Cartan-Altland-Zirnbauer scheme for topological superconductors~\cite{Altland1997Nonstandard,Sato2017Topological}.

 To access the effective MZM and YSR energies, we express the Hamiltonian in Eq.~\eqref{eq:EffectiveH} in a full fermionic basis by
 $f=\frac{1}{2}(\gamma_{\rm L}^{} + i \gamma_{\rm R}^{})$,
 where $f^\dagger$ is the creation operator of the non-local fermion with energy $\epsilon_{\rm M}^{}$. The four single-particle eigenvalues are,
\begin{equation}\label{eq:EffectiveE}
\begin{split}
 E_{\rm{S}}^{} & = \pm \tfrac{1}{2}\left[ \sqrt{\epsilon_{+}^2+4|t_{+}|^2} + \sqrt{\epsilon_{-}^2+4|t_{-}|^2}\right]
\\
 E_{\rm{M}}^{} & = \pm \tfrac{1}{2}\left[ \sqrt{\epsilon_{+}^2+4|t_{+}|^2} -  \sqrt{\epsilon_{-}^2+4|t_{-}|^2}\right],
 \end{split}
\end{equation}
where
$t_{\pm}= t_{\rm L}^{} \pm i t_{\rm R}^{}$ 
and
$\epsilon_{\pm}=\epsilon_{\rm M}^{}\pm\epsilon_{\rm S}^{}$.
The two solutions with larger energy value, $|E_{\rm S}^{}|$, are the effective YSR energies, while the two solutions with smaller energy value, $|E_{\rm M}|$,  are the effective hybridization energies of MZM. The localization of the corresponding states of $E_{\rm M/S}^{}$  strongly depends on the parameters of the single magnetic adatom; see Appendix~\ref{app:LDOS} for details.
The effective energies are unchanged under the adiabatic exchange of
$\gamma_{\rm L}^{}$ and $\gamma_{\rm R}^{}$,
which leads to 
$\gamma_{\rm L}^{} \rightarrow -\gamma_{\rm R}^{}$ 
and
$\gamma_{\rm R}^{} \rightarrow \gamma_{\rm L}^{}$~\cite{Leijnse2012Introduction}.
This implies 
$\epsilon_{\rm M}\rightarrow \epsilon_{\rm M},\, t_{\rm L}^{}\rightarrow t_{\rm R}^{},\, t_{\rm R}^{}\rightarrow-t_{\rm L}^{}$,
leaving Eq.~\eqref{eq:EffectiveE} invariant. Also, swapping the labels R and L in Eq.~\eqref{eq:EffectiveH}, leaves Eq.~\eqref{eq:EffectiveE} unchanged since
$\epsilon_{\rm M}\rightarrow -\epsilon_{\rm M} \implies \epsilon_{\pm} \rightarrow-\epsilon_{\pm}$,
and all other parameters are unaffected. 

We are most interested in a vanishing effective Majorana hybridization, $E_{\rm M}$. Importantly, in the symmetry class $D$, this corresponding zero-energy crossing of the Majorana level is generally accompanied by a change in the fermion ground state parity, which is conveniently described by $P= \sgn\left(\rm Pf\left[\tilde{H}_{\rm eff}\right]\right)$, i.e., the sign of the Pfaffian of $\tilde{H}_{\rm eff}$, which is obtained by writing $H_{\rm eff}$ in Majorana basis~\cite{kitaev2001unpaired,nadj2013proposal,cheng2021fatemajoranazeromodes}. 
We hence find that the effective Majorana hybridization crosses zero and the ground-state parity changes only at
 \begin{equation}\label{eq:EMeq0}
     \Im[t_{\rm R}] \Re[t_{\rm L}]-\Im[t_{\rm L}]\Re[t_{\rm R}]=\dfrac{\epsilon_{\rm M}^{}\epsilon_{\rm S}^{}}{4},
 \end{equation}
 see Appendix~\ref{app:Parity} for details.
 One scenario to satisfy Eq.~(\ref{eq:EMeq0}) is when one of $t_{\rm L}^{}$ and $t_{\rm R}^{}$ is zero or both are real (both of which give $|t_-|=|t_+|$) and $\epsilon_{\rm S}^{}$ or $\epsilon_{\rm M}$ is zero. The investigation of more general possibilities, when both sides of Eq.~\eqref{eq:EMeq0} are finite, requires knowledge about the parameters in Eq.~\eqref{eq:EMeq0}, which depend on the microscopic model; see inset in Fig.~\ref{fig:Fig1}(a) for a representative dependence of the effective Majorana hybridization on the magnetic orientation of the single adatom.
An important implication of the above result is that parity switching, which can be determined through charge sensing~\cite{Burnell2013Measuring,Schulenborg2020Parity,Tsintzis2022Creating,Nitsch2022Interference}, can serve as a tool for tracking the MZMs shielding.  Furthermore, controlling ground-state parity is essential in parity-based measurements of Majorana correlations~\cite{Burnell2013Measuring}. Here, one can perturb one of the chains or control its parity through the parameters of the single adatom while simultaneously measuring the parity of the second chain, which is expected to change as a result of the entanglement between the MZMs.

While the control of the Majorana hybridization and the accompanying parity switching are general for the Majorana setups in Fig.~\ref{fig:Fig1}, the microscopic details of the systems are required in order to examine the dependence of the MZMs hybridization on the control parameters of the YSR state. We explore the experimentally relevant realizations in Fig.~\ref{fig:Fig1} for this purpose. 
In what follows, we, therefore, investigate the widely studied arrangements of magnetic adatoms on superconductors with helical magnetization and the ones with ferromagnetic ordering. 
%
\section{Physical realizations}\label{sec:Control} 
In this Section, we describe in detail physical realizations of magnetic chain-superconductor systems and show explicitly the dependence of the hybridization of MZMs on the control parameters. In Sec.~\ref{subsec:Cont}, we consider continuum models for superconducting substrates, derive general expressions for $t_{\rm L/R}$, and subsequently demonstrate the control of Majorana hybridization for helical and ferromagnetic chains. In Sec.~\ref{sec:LatticeModel}, we consider ferromagnetic chains on a lattice superconductor and show that the model yields the same results as the former, demonstrating the universality of our findings.
%
\subsection{Continuum models for the substrate superconductor} \label{subsec:Cont} 
We first discuss models where the superconducting substrate is incorporated by continuous fields. In the dilute adatom limit, $k_{ \rm F}a \gg 1$, where $k_{\rm F}$ is the Fermi wave number of the superconducting substrate, an effective Hamiltonian that captures the basic physics of this setup can be written in the basis of the YSR states induced by the individual adatoms in the chain if the energy of each YSR state is close to the Fermi energy \cite{Pientka2013topological,brydon2014topological}. Following Refs.~\cite{Pientka2013topological,brydon2014topological}, we apply the approach to the setup in Fig.~\ref{fig:Fig1}(b) and obtain the effective Hamiltonian as, 
\begin{widetext}
 \begin{equation}\label{eq:GenContH}
     \begin{split}
& \mathcal{H}_{}=\mathcal{H}_{\rm L} + \mathcal{H}_{\rm R} +  \mathcal{H}_{\rm LR}  + \mathcal{H}_{\rm d} \\
 & \mathcal{H}_{\rm L} =\sum_{m=1}^{N_{L}} \sum_{n=1}^{N_{L}} b_m^\dagger\left[\epsilon_{0}\delta_{mn} + \left(1-\delta_{mn} \right)h_{mn}^{} \right] b_n + b_m \left(1-\delta_{mn} \right) \Delta_{mn} b_n + {\rm h.c.},\\
& \mathcal{H}_{\rm R}  =\sum_{m=N_L+R/a}^{N_{L}+N_R+R/a} \sum_{n=N_{L}+R/a}^{N_{L}+N_R+R/a} b_m^\dagger\left[\epsilon_{0}\delta_{mn} + \left(1-\delta_{mn} \right)h_{mn}^{} \right] b_n + b_m \left(1-\delta_{mn} \right) \Delta_{mn} b_n + {\rm h.c.},\\
& \mathcal{H}_{\rm LR} =\sum_{m=1}^{N_L} \sum_{n=N_{L}+R/a}^{N_{L}+N_{R}+R/a} \left[ h_{mn}^{}  b_m^\dagger b_n + b_m \left(1-\delta_{mn} \right) \Delta_{mn} b_n \right]  + {\rm h.c.},\\
&  \mathcal{H}_{\rm S} = \epsilon_{\rm S}^{} d_{}^\dagger d_{} + \left[\sum_{m=1}^{N_{L}}\left(h_{N_{L}+\frac{R_{\rm S}^{}}{a},m}^{} d_{}^\dagger + \Delta_{N_{L}+\frac{R_{\rm S}^{}}{a},m} d_{}\right) + \sum_{m=N_L+R/a}^{N_L+N_R+R/a}  \left( h_{N_{L}+\frac{R_{\rm S}^{}}{a},m}^{} d_{}^\dagger  
  + \Delta_{N_{L}+\frac{R_{\rm S}^{}}{a},m}^{} d_{} \right)\right] b_{m} + {\rm h.c.}, 
\end{split}
 \end{equation}
\end{widetext}
where $b_m^\dagger$ creates an electron with energy $\epsilon_0$ in the YSR state at site $m=\left(x_m,y_m,z_m\right)$ in the chain and $d^\dagger$ creates an electron with energy $\epsilon_{\rm S}$ in the YSR state at the position of the single magnetic adatom. Here, $\mathcal{H}_{\rm LR}$ couples the YSR states in the left chain to those in the right chain and vice versa while the second term of $\mathcal{H}_{\rm S}$ couples the YSR state from the single magnetic adatom to all YSRs states in the left and right chains. In the general form, the long-range hopping $h_{m n}$ and order parameters $\Delta_{m n}$ are given by~\cite{brydon2014topological,Schneider2022Testing},
\begin{equation}\label{eq:GenContMatrixElement}
\begin{aligned}
h_{m n} & =h_{m n}^{(0)}\langle\uparrow m \mid \uparrow n\rangle+h_{m n}^{(1)}\left\langle\uparrow m\left|i \sigma_y\right| \uparrow n\right\rangle, \\
\Delta_{m n} & =\Delta_{m n}^{(0)}\langle\uparrow m \mid \downarrow n\rangle+\Delta_{m n}^{(1)}\left\langle\uparrow m\left|i \sigma_y\right| \downarrow n\right\rangle,
\end{aligned}
\end{equation}
where $\sigma_\nu$ is the $\nu$-Pauli matrix, and $h_{m n}^{(1)}$ and $\Delta_{m n}^{(1)}$ are material-dependent constants that vanish when spin-orbit coupling is absent in the superconductor. Here,
\begin{equation*}
\begin{aligned}
 & |\uparrow m\rangle =\left(
\cos \left( \theta_{ m}^{} / 2\right) e^{-i \phi_{ m}^{} / 2} ,
\sin \left(\theta_{ m}^{} / 2\right) e^{i \phi_{ m}^{} / 2}
\right)^T,\\
&|\downarrow m\rangle = \left( \sin \left(\theta_{ m}^{} / 2\right) e^{-i \phi_{ m}^{} / 2} , 
-\cos \left(\theta_{ m}^{} / 2\right) e^{i \phi_{ m}^{} / 2}\right)^T,
\end{aligned}
\end{equation*}
are the spinors corresponding to the magnetic orientations of the adatoms in the chains at position $\bs r_m$, which is denoted by $ S\left(\sin \theta_{ m}^{} \cos\phi_{ m}^{},\sin \theta_{ m}^{} \sin\phi_{m}^{},\cos \theta_{ m}^{} \right)$, where $S$ is the magnitude of the spin and $\theta_{ m}^{}$ and $\phi_{ m}^{}$ are the polar and azimuthal angles, respectively~\cite{Pientka2013topological,braunecker2013interplay,brydon2014topological}.
Due to the phase factors in $h_{mn}$ and $\Delta_{m n}$ in Eq.~\eqref{eq:GenContMatrixElement}, $\mathcal{H}_{\rm L}$ and $\mathcal{H}_{\rm R}$ in Eq.~\eqref{eq:GenContH} break time-reversal symmetry but preserve particle-hole symmetry, thus, the chains are class $D$ topological superconductors~\cite{Altland1997Nonstandard,Kitaev2009,Ryu2010,Sato2017Topological} like the effective low-energy model in Eq.~\eqref{eq:EffectiveH}.

While the bare Majorana hybridization, $\epsilon_{\rm M}$, depends only on the distance R between the chains and the material-specific details, the low-energy coupling $t_{\rm L/R}$ on the other hand depends also on the parameters associated with the single magnetic adatom. We find that $t_{\rm L/R}$ have separable dependencies on the magnetic adatom's spin orientation and the remaining parameters. By using Eqs.~\eqref{eq:GenContH} and~\eqref{eq:GenContMatrixElement} with Eq.~\eqref{eq:EffectiveH}, (see Appendix~\ref{app:tGen} for details) we obtain the general form of the couplings,
\begin{equation}\label{eq:tGen}
 t_{\rm L/R} =e^{i\frac{\phi_{\rm S}^{}}{2}}\cos\left(\frac{\theta_{\rm S}^{}}{2}\right)\, F_{L/R} + e^{-i\frac{\phi_{\rm S}^{}}{2}}\sin\left(\frac{\theta_{\rm S}^{}}{2}\right)\, G_{\rm L/R},
\end{equation}
where the functions 
$G_{\rm L/R}^{}=G_{\rm L/R}^{}\left(R_{\rm S}^{}\right)$ and $F_{\rm L/R}=F_{\rm L/R}^{}\left(R_{\rm S}^{}\right)$
contain the microscopic details of the superconductor and magnetic chain. Here, $F_{L/R}$ can be obtained from $G_{L/R}^{}$ by flipping the direction of every spin of each chain, $
    F_{L/R}^{}(\phi_{ m}^{},\theta_m)= G_{L/R}^{}(-\phi_{ m},\pi-\theta_m)$. Equation~\eqref{eq:tGen} is valid for all magnetic adatoms chain-superconductor systems in the dilute and deep Shiba limits.

Next, we present two specific continuum models of our setup and demonstrate the control of the Majorana hybridization.
%
%
\subsubsection{Helically magnetized chains on a conventional $s$-wave superconductor} \label{subsubsec:PientkaModel} 
A well-studied model is the helical arrangement of the magnetic adatoms chain on the SC due to the RKKY interactions~\cite{menzel2012information,Pientka2013topological,li2016manipulating}. Such a spin structure was found for iron adatoms on the surface of superconducting rhenium~\cite{howon2018Toward}.  

In the absence of spin-orbit coupling, corresponding to $h_{m n}^{(1)}=\Delta_{m n}^{(1)}=0$ in Eq.~\eqref{eq:GenContMatrixElement}, analytical zero energy Majorana solutions of $\mathcal{H}_{\rm L}$ and $\mathcal{H}_{\rm R}^{}$ can be found in the vicinity of the Bragg point,  $k_h = k_{\rm F}$, where $k_h=2\pi\chi$ is the wave vector corresponding to the periodicity of the helix $\chi$~\cite{Pientka2014Unconventional}. We first focus on the Bragg point for building up a physical understanding. We have explored deviations from this point numerically and found no qualitative difference when the system remains in the topologically nontrivial phase. Following Ref.~\cite{Pientka2013topological}, we consider an in-plane helical magnetization in the chains, which corresponds to setting $\theta_m=\pi/2$. We further specify the helical structure for the left (right) chain as $\phi_{{\rm L},m}=-2k_ham$ $\left(\phi_{{\rm R},m}=2k_ham\right)$, and we fix $m=0$ at the inner ends of the chains, which are both oriented in the $x$~direction.
For analytical tractability, we consider the semi-infinite limit of both chains such that the outer MZMs, $\gamma_{\rm L/R}^\prime$, are at infinity. 
For this case, we find that $G_{\rm L}(R_{\rm S}^{})=-iF_{\rm L}^*(R_{\rm S}^{})$ and $G_{\rm R}(R_{\rm S}^{})=iF_{\rm R}^*(R_{\rm S}^{})$, see Appendix~\ref{SpecialCaseAppendix}. Using this result and the Majorana solutions of $\mathcal{H}_{\rm L}$ and $\mathcal{H}_{\rm R}$~\cite{Pientka2014Unconventional}, we find the effective low-energy parameters in Eq.~\eqref{eq:EffectiveH}, see Appendix~\ref{app:ToyModelParams}, as
\begin{equation}\label{eq:LowEPientkaPar}
\begin{split}
& \epsilon_{\rm M}=\frac{ i B^2}{ \Delta k_{\rm F} a}\sum_{\substack{w=1,2\\ \zeta=\pm}}^{} \zeta e^{-A_\zeta} \frac{{}_2F_1\left(w, \frac{R}{ a}+w-1 ; \frac{R}{ a}+w ; e^{A_\zeta}\right)}{\frac{R}{ a}+w-1},\\
& t_{\rm L}= t_{\rm L}\left(F_{\rm L}\rightarrow G,G_{\rm L} \rightarrow -i G^*\right),\\
& t_{\rm R}\left(R_{\rm S}^{},\phi_{\rm S}^{},\theta_{\rm S}^{}\right) = t_{\rm L}^*\left( R-R_{\rm S}^{},-\phi_{\rm S}^{},\theta_{\rm S}^{}\right),  \\ 
& G  = \frac{Be^{-\left[ik_{\rm F} R_{\rm S}^{}+ \frac{R_{\rm S}^{}}{\xi_{\rm sc}} + i \frac{\pi}{4}\right]}}{2 k_{\rm F} R_{\rm S}} \ {}_2F_1\left(1, \frac{R_{\rm S}^{}}{a}; \frac{R_{\rm S}^{}}{a}+1;e^{A_{-}}\right).
\end{split}
\end{equation}
Here, $B^2=\frac{ \left(1-\beta^2 \right) \Delta^2}{2}$, 
$\beta=\frac{e^{a / \xi_{\rm sc}} \sin \left(k_F a \epsilon_0 / \Delta\right)}{\sin \left(2 k_F a+k_F a \epsilon_0 / \Delta\right)}$, $A_\zeta= -\frac{a}{\xi_{\rm sc}} + \ln{(\beta)+i\zeta 2k_{\rm F}a}$, ${ }_2 F_1$ is the ordinary hypergeometric function and $\xi_{\rm sc}$ is the superconducting coherence length of the substrate. For the effective model parameters in Eq.~(\ref{eq:LowEPientkaPar}), we use a 3D bulk superconductor in the model and also take $\bs R_{\rm S}^{} = \left(x_{\rm S}^{},\, 0,\, 0\right)$ for analytic convenience. If the inter-chain distance $R$ is large or the single magnetic adatom is far from the chains, the effective model parameters are small and can be represented by their asymptotic form, see Appendix~\ref{app:ShieldingAsymp}.
%
%
\begin{figure*}[!t]
 \includegraphics[width=1.0\textwidth]{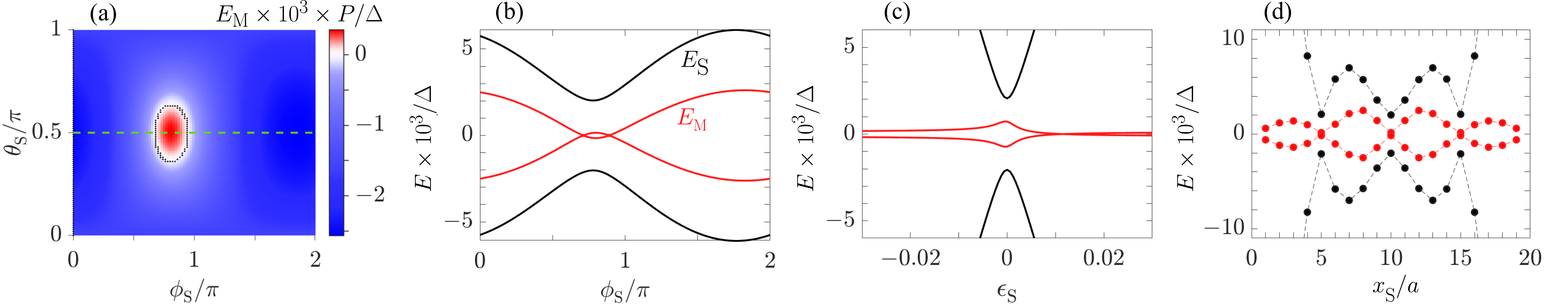}
	\caption{Majorana hybridization at the Bragg point of a chain of helically-oriented planar spins on an $s$-wave superconductor (continuum model). 
 (a) The effective Majorana energy multiplied with the ground state parity, $E_{\rm M}\times P$, as a function of the orientation of the single magnetic adatom $\phi_{\rm S}^{}$ and $\theta_{\rm S}^{}$ for bare adatom energy $\epsilon_{\rm S}^{}=0.002\Delta$ and the position relative to the right edge of the left chain $\bs R_{\rm S}^{}= \left(8a,\, 0,\, 0\right)$. The red and blue regions have opposite parity, while the MZMs are decoupled in between. The black line denotes the exact decoupling parameters, $E_{\rm M}=0$.
 (b) Linecut along the dashed green line in (a). 
 (c) Hybridization as a function of the YSR energy $\epsilon_{\rm S}^{}$ at $\phi_{\rm S}^{}=\pi$, and 
 (d) as a function of the distance from the left chain $R_{\rm S}^{}$ at $\phi_{\rm S}^{}=0$. Here, $\bs R_{\rm S}^{} = \left(x_{\rm S}^{},\, 0,\, 0\right)=x_{\rm S}^{}$ is plotted in steps of the inter magnetic adatom distance, $a$, in the chains. In all panels, we set the distance between the chains $R=20a$, the bare energy of the magnetic adatoms of the chains $\epsilon_{\rm 0}^{}=0.005\Delta$, and the superconducting coherence length $\xi_{\rm sc}=15a$.}
\label{fig:Fig2}
\end{figure*}

Combining Eqs.~\eqref{eq:LowEPientkaPar} and~\eqref{eq:EffectiveE}, we next track both the shielding effect and the tuning of the Majorana hybridization. First, we focus on controlling the Majorana hybridization and its shielding with the single adatom's magnetic orientation, which is the parameter that can be altered fastest in an experiment, being adaptable without changing the positions of the adatoms. Notably, the effective Majorana hybridization, $E_{\rm M}$, is tuned through a wide range of values, see Fig.~\ref{fig:Fig2}(a), where the blue and the red region denote parameter regimes of different ground state parity. Remarkably, at the boundary between these parameter regions, the MZMs are completely decoupled, $E_{\rm M}=0$, as described by Eq.~\eqref{eq:EMeq0}. Such a parity switch, seen in the change in the sign of $E_{\rm M}$, around this region confirms the zero-energy crossing of the MZMs. 
%

To further understand the Majorana shielding effect for the model, we consider the case where $\theta_{\rm S}^{}=\pi/2$, see dashed green line in Fig.~\ref{fig:Fig2}(a). In this case, Eq.~\eqref{eq:tGen} and Eq.~\eqref{eq:LowEPientkaPar} imply $\Im[t_{\rm L}^{}]=-\Re[t_{\rm L}^{}]$ and $\Im[t_{\rm R}^{}]=\Re[t_{\rm R}^{}]$. Substituting these conditions in the general shielding condition in Eq.~\eqref{eq:EMeq0}, we obtain $\Re[t_{\rm R}^{}] \Re[t_{\rm L}^{}]={\epsilon_{\rm S}^{}\epsilon_{\rm M}}/{8}.$ In the limit of the realistic condition $\epsilon_{\rm S}^{} \epsilon_{\rm M}^{} \approx 0$, the condition for the crossings at $\theta_{\rm S}^{}=\pi/2$ reduces to $E_{\rm M}^{}=0 \iff \Re[t_{\rm R}^{}] \Re[t_{\rm L}^{}]\approx0$. The two distinct crossings of $E_{\rm M}$ can be seen in Fig.~\ref{fig:Fig2}(b). Figure~\ref{fig:Fig2}(b) also shows that $E_{\rm S}^{}$ has a similar angular dependence as $E_{\rm M}$, but remains finite. 
%

The above result motivates us to find expressions for the zeros of $t_{\rm L}^{}$ and $t_{\rm R}^{}$. For this purpose, we determine the magnetic orientations of the single magnetic adatom $\phi_{S,L}(R_{\rm S}^{})$, $[\phi_{\rm S,R}^{}(R_{\rm S}^{})]$ for which $t_{\rm L}^{}=0$, $[t_{\rm R}^{}=0]$, using Eq.~\eqref{eq:LowEPientkaPar}.   We obtain the expression (see Appendix~\ref{app:ShieldingAsymp} for asymptotic forms),
\begin{equation}\label{ZeroPhi}
    \begin{split}
        \phi_{\rm S,L}^{}(R_{\rm S}^{})=&-2k_F R_{\rm S}^{} +
 2  \tan^{-1}\left(\varphi\right)\mod 2\pi, \\
 \phi_{\rm S,R}^{}(R_{\rm S}^{})=&-\phi_{\rm S,L}^{}\left(R-R_{\rm S}^{}\right),
    \end{split}
\end{equation}
where $\varphi=\arg \left[_2F_1(1,R_{\rm S}^{}/a,R_{\rm S}^{}/a+1,e^{-A_-})\right]$ and $A_-$ is defined below Eq.~(\ref{eq:LowEPientkaPar}). In general, for every position of the single magnetic adatom, Eq.~\eqref{ZeroPhi} provides distinct angles, $\phi_{\rm S, L/R}^{}$ for $\theta = \pi/2$, for which $t_{\rm L}^{}$ and $t_{\rm R}^{}$ vanish.

%
%
The bare YSR energy, $\epsilon_{\rm S}^{}$, and the distance of the magnetic adatoms from the chains can also be used to control the Majorana hybridization $E_{\rm M}$, as shown in Fig.~\ref{fig:Fig2}(c) and (d), respectively, demonstrating the flexibility in controlling $E_{\rm M}$. As seen in Fig.~\ref{fig:Fig2}(d), the effective YSR energy $E_{\rm S}^{}$ oscillates symmetrically about  $R_{\rm S}^{}=R/2$. The wavelength of this oscillation is determined by $k_F$, such that the hybridization oscillates multiple times within the lattice constant $a$ in the dilute impurity limit, which we do not depict assuming the magnetic impurity is placed at defined positions on the lattice only. We note that a larger value for the coherence length $\xi_{\rm SC}$ than the choice we made for Fig.~\ref{fig:Fig2} increases the tunability of $E_{\rm M}$. This is because the coupling parameters $t_{\rm L/R}$ are suppressed for distances much larger than the coherence length, as shown in Eq.~\eqref{eq:LowEPientkaPar}. This highlights the applicability of the shielding effect beyond our setup. When the single magnetic adatom is close to one of the magnetic chains, $t_{\rm L}\approx 0$ or $t_{\rm R}\approx 0$ and $E_{\rm S}^{}$ has the largest deviation from $\epsilon_{\rm S}^{}$ due to strong overlap of the YSR state with one of the MZMs. Consistently, for the case when the adatom is close to the left chain, if additionally the bare Majorana coupling vanishes ($\epsilon_{\rm M}^{}=0$), then $E_{\rm S}^{}=\pm \sqrt{\epsilon_{\rm S}^2+4|t_{\rm L}|^2}$ and $E_{\rm M}^{}=0$, which implies that the system behaves like a single magnetic chain with a single additional magnetic adatom close by.

For the Bragg point, see Fig.~\ref{fig:Fig2}, we find a good agreement between analytical and numerical results when $E_{\rm M}^{},\,E_{\rm S}^{} \ll \Delta_{\rm t}$ which marks the regime where the higher-energetic Shiba bands can be safely neglected. Away from this limit, while both analytical and numerical results agree qualitatively there is a noticeable quantitative difference as expected, especially when the YSR state couples very strongly to one of the MZMs (see Appendix~\ref{app:Error} for more details). The numerical results are obtained by diagonalizing Eq.~\eqref{eq:GenContH}, for $h_{m n}^{(1)}=\Delta_{m n}^{(1)}=0$ in Eq.~\eqref{eq:GenContMatrixElement}, within the Bogoliubov-de-Gennes (BdG) formalism with $N_{\rm L}=N_{\rm R}=100$. For this length, we confirmed that the outer MZMs have negligible overlap with inner MZMs or the single YSR state. While our analytical evaluation considers the single magnetic adatom to be on the $x$~axis, i.e., aligned with the chains, we have checked numerically that the magnetic adatom can still be used to tune $E_{\rm M}^{}$ even if the adatom is displaced from the $x$~axis as long as the MZMs and the YSR state overlap is sufficiently strong. 
%
%
 \begin{figure*}[!t]
\includegraphics[width=0.95\textwidth]{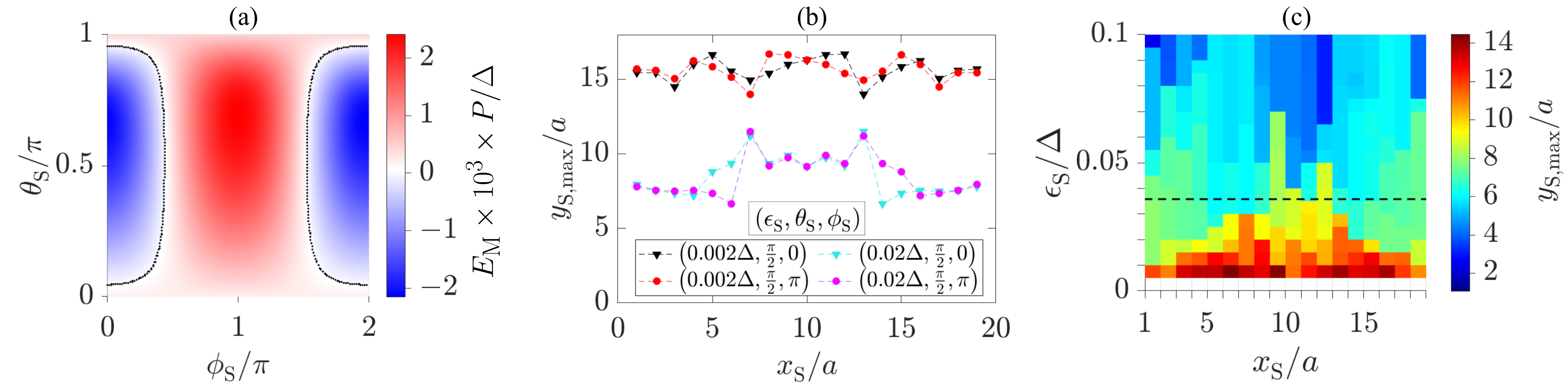}
\caption{Majorana hybridization for a ferromagnetic chain on a 2D spin-orbit coupled superconductor (continuum model):
 (a) Effective Majorana hybridization $E_{\rm M}^{}\times P$ as a function of the magnetic orientation of the single magnetic adatom $\theta_{\rm S}$ and $\phi_{\rm S}$ for $\epsilon_{\rm S}=0.002\Delta,~\bs R_{\rm S}^{}=(8a,0.58a)$. The black line denotes the exact decoupling parameters, $E_{\rm M}=0$. This result is qualitatively the same as Fig.~\ref{fig:Fig2}(a).
 (b) Maximal distance $y_{\rm S, max}$  of the adatom to the $x$~axis where decoupling can be achieved in dependence on $x_{\rm S}^{}$ for several $\phi_{\rm S}$ and $\epsilon_{\rm S}^{}$ values and fixed $\theta_{\rm S} = \pi/2$.
 (c) Maximal distance $y_{\rm S, max}$  of the adatom to the $x$~axis where decoupling can be achieved as a function of $x_{\rm S}^{}$ and $\epsilon_{\rm S}^{}$ for fixed $\theta_{\rm S} = \pi/2,\,\phi_{\rm S} = 0$.
 In all panels, $\xi_{\rm sc}=5a,\,R=20a,\,N_{\rm L}=N_{\rm R}=100,\,\epsilon_0=0.005\Delta$, and spin-orbit coupling strength $\lambda=0.09\Delta$. The value of the topological gap is $\Delta_{\rm t}\approx0.04\Delta$ (black dashed line in Fig. (c)).} 
\label{fig:Fig3} 
\end{figure*}
\subsubsection{Ferromagnetic chains on an $s$-wave superconductor with nonvanishing spin-orbit coupling}\label{subsubsec:BrydonModel}
In this section, we turn our attention to ferromagnetic chains on a superconductor with spin-orbit coupling, see Fig.~\ref{fig:Fig1}(a). The model describes for example Fe adatoms on a superconducting Pb~\cite{nadj2014observation} and Mn adatoms on Nb superconductor~\cite{Schneider2022}. Following Ref.~\cite{brydon2014topological}, we consider a 2D continuum superconductor and extract the matrix elements for both the left and right chains according to Eq.~\eqref{eq:GenContMatrixElement}, see Appendix~\ref{app:FMeffective} for details. Here, an analytic solution to the Majorana wavefunction has, to the best of our knowledge, not yet been found because of the complicated form of the matrix elements~\cite{brydon2014topological}. We, therefore, proceed numerically by diagonalizing Eq.~\eqref{eq:GenContH} in the BdG formalism for a chain of $N_{\rm L}=N_{\rm R}=100$, in the topologically nontrivial regime. 

The Majorana hybridization is controllable like in the helical chain model in Sec.~\ref{subsubsec:PientkaModel}, and we find that the ground state parity switches when crossing $E_{\rm M}=0$ as the orientation of the single magnetic adatom varies, see Fig.~\ref{fig:Fig3}(a). Comparing Fig.~\ref{fig:Fig3}(a) and Fig.~\ref{fig:Fig2}(a), we see that the results from both physical realizations qualitatively agree, and the  MZM hybridization can be tuned to zero energy by fine-tuning the parameters. When the single magnetic adatom is placed at $R_{\rm S} =R/2$, the effective Majorana energy $E_{\rm M}$ does not depend on the azimuthal angle $\phi_{\rm S}$ of the single adatom's magnetization, see Appendix~\ref{app:GenFM} for a proof.

To determine the threshold overlap between the MZMs and the YSR wavefunction that is necessary to control $E_{\rm M}$, we numerically search for the farthest position of the single magnetic adatom from the chains, $y_{\rm S,max}^{}$, beyond which Majorana shielding is no longer possible, i.e.,  Eq.~\eqref{eq:EMeq0} is not satisfied, and there is no change in parity. As expected,  $y_{\rm S,max}^{}$ depends on the single magnetic adatom parameters, see Fig.~\ref{fig:Fig3}(b,c). It is noteworthy  that the Majorana hybridization can be tuned to zero for $\epsilon_{\rm S}^{}$ values larger than the topological gap, see Fig.~\ref{fig:Fig3}(c). With increasing $\epsilon_{\rm S}^{}$, the bare YSR energy moves away from $\epsilon_{\rm M}^{}$,  which reduces the coupling of the YSR state to the MZMs leading to the decrease in $y_{\rm S, max}^{}$. Similarly, positioning the magnetic adatom closer to one of the magnetic chains reduces the overlap of the YSR state with the MZM on the other chain, which further decreases with increasing $y_{\rm S}^{}$. This leads to a smaller $y_{\rm S, max}^{}$ compared to an adatom present near the middle of the two chains. The values are not symmetric about $x_{\rm S}^{}=R/2$ due to the absence of mirror symmetry for a fixed magnetic orientation of the single magnetic adatom. However, mirroring the setup and simultaneously flipping the spin of the single magnetic adatom is a valid symmetry, which results in $y_{\rm S,max}^{}(x_{\rm S}^{},\phi_{\rm S}=0)=y_{\rm S,max}^{}(R/2-x_{\rm S}^{},\phi_{\rm S}=\pi)$. To achieve full control over the Majorana hybridization, including the complete shielding of Majorana hybridization and the change of the ground state parity, $y_{\rm  S}$ should be kept within its maximum limit. This condition applies to all magnetic chain-superconductor setups.

Having demonstrated the Majorana shielding effect with continuum superconductors we next turn our attention towards lattice models for the superconducting substrate.
%
\subsection{Lattice model for the superconducting substrate: Ferromagnetic chains on a  superconductor with non-vanishing spin-orbit coupling}\label{sec:LatticeModel}
%
\begin{figure*}[!t]
\includegraphics[width=0.95\textwidth]{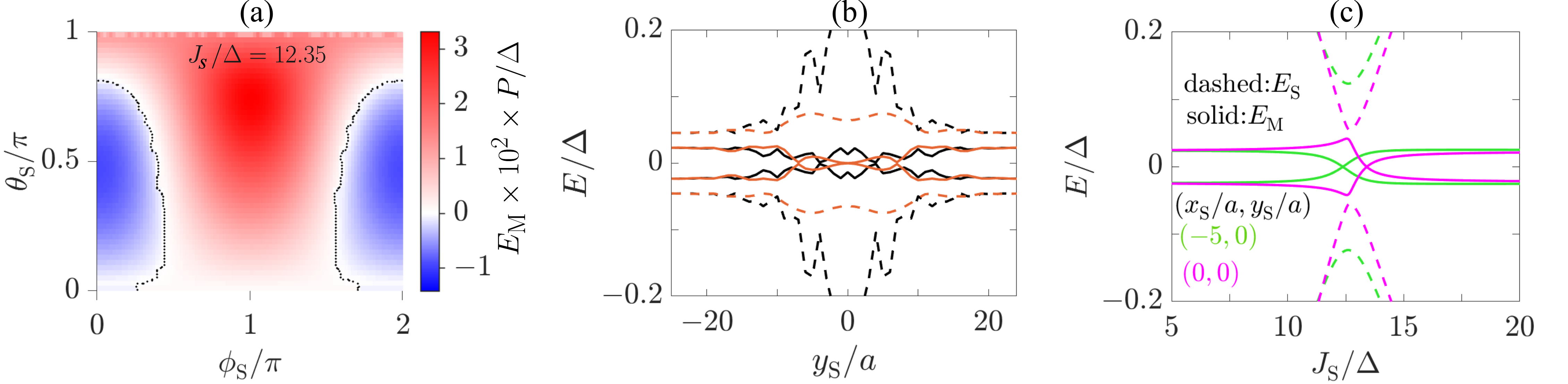}
\caption{Majorana hybridization for ferromagnetic chains on a 2D square lattice $s$-wave superconductor with spin-orbit coupling.
 (a) Effective Majorana hybridization $E_{\rm M}^{}\times P$ as a function of $\phi_{\rm S}^{}$ and $\theta_{\rm S}^{}$ for a single magnetic adatom placed at $(x_{\rm S}^{}/a,y_{\rm S}^{}/a)=(-5,0)$. The black line denotes the exact decoupling parameters, $E_{\rm M}=0$. The behavior is qualitatively similar to continuum superconductors shown in Figs.~\ref{fig:Fig2}(a) and Fig.~\ref{fig:Fig3}(a). 
 (b)  Effective YSR and Majorana hybridization $E$ vs. $y$ for a single magnetic adatom with fixed $x_{\rm S}^{}/a=-5$ (black) and $x_{\rm S}^{}/a=5$ (brown) for $\left(\phi_{\rm S}^{},\theta_{\rm S}^{}\right)=\left(0,0\right)$. 
 (c) $E$ vs. $J_{\rm S}^{}$ for a magnetic adatom at different positions. The calculation is carried out on an $81\times 81$ square lattice with $\Delta=0.2t,\, \lambda =0.3t,\, \mu = 0.5t,\, t=1$ and $R/a=21$. The chains are in the topologically nontrivial phase with $J/\Delta= 11.2$, resulting in $\epsilon_{\rm M}^{}/\Delta =0.035,\, \Delta_{\rm t}/\Delta\approx 0.4$.}
\label{fig:Fig4}
\end{figure*}
We next consider ferromagnetic adatom chains on a 2D superconducting square lattice with Rashba spin-orbit coupling, which has been used to describe several experimental reports of signatures of topological superconductivity in atomic chains~\cite{nadj2014observation,ruby2015end,schneider2021topological,Schneider2022}. The lattice model incorporates the full bandwidth and the lattice symmetries of the system. The continuum models in Sec.~\ref{subsec:Cont} can be obtained by expanding the lattice model around small momenta and keeping only the lowest orders.

In order to a priori exclude the outer MZMs, $\gamma_L^\prime$ and $\gamma_R^\prime$, depicted in Fig.~\ref{fig:Fig1}(a), we consider a slightly altered setup, where the outer ends of the chains are connected by periodic boundary conditions. We similarly use periodic boundary conditions for the substrate.
Thus, the components of the Hamiltonian are~\cite{Bjornson2016Majorana, Awoga2017Disorder, theiler.bjornson19, mashkoori2020identification}
\begin{equation}\label{eq:LattModel}
   \begin{split}
  & H = H_{\rm SC}^{} + H_{\rm R+L}^{} + H_{\rm S}^{} \text{, where}\\
    &	H_{\rm SC}^{}  =  \sum_{\mathbf  m, \mathbf b, \alpha} \left[ \left(4t-\mu \right) c^\dagger_{\mathbf m,\alpha} c_{\mathbf m,\alpha}^{} - t c^\dagger_{\mathbf m+ \mathbf b,\alpha} c_{\mathbf m  ,\alpha}^{} \right] \\
	&	+ \sum_{\mathbf m, \mathbf b} \left[   - \lambda_{\rm } e^{i\theta_{\mathbf b}} c_{\mathbf m + \mathbf b,\downarrow}^\dagger c_{\mathbf m,\uparrow} + \Delta  c^\dagger_{\mathbf m,\uparrow}c^\dagger_{\mathbf m,\downarrow} +\mathrm{h.c.}\right] ,\\
	&	H_{\rm R+L}^{}  =  J S \sum_{ m_y}^{}\sum_{ m_x =1}^{N_{\rm L}+N_{\rm R}} \sum_{\alpha,\beta}\sigma_{\alpha\beta}^z c_{\mathbf m,\alpha}^\dagger c_{\mathbf m,\beta}^{}\,\delta_{m_y,\frac{M_y+1}{2}}^{},\\
	&	H_{\rm S}^{}  = J_{\rm S}^{}\sum_{\alpha,\beta}(\bs\sigma\cdot\bs S_{\rm S})_{\alpha\beta} c_{N_{\rm L} +R_{\rm S}^{},\alpha}^\dagger c_{N_{\rm L}+R_{\rm S}^{},\beta}^{}.
\end{split}
\end{equation}
 Here, $H_{\rm R+L}$ is the Hamiltonian of the left and the right chains, connected by the periodic boundary condition. The operator $c^\dagger_{m,\alpha}$ creates an electron with spin $\alpha$ at site $\mathbf m=\left(m_x,m_y\right)$ on the $M_x\times M_y$ 2D lattice, while $\sigma^{\nu}$ are the Pauli matrices in spin space. In $H_{\rm SC}$, $t$ is the nearest neighbor hopping strength, $\mu$ is the homogeneous chemical potential, and $\lambda_{\rm }$ is the strength of the Rashba spin-orbit coupling, which has directional dependencies given by the polar coordinates $\theta_{\mathbf b}$ of the nearest-neighbor bond vectors $\mathbf b$.  This Hamiltonian, in addition to breaking time-reversal symmetry through $H_{\rm R+L}$ and $H_{\rm S}$, also breaks the spin rotation symmetry through spin-orbit coupling. Thus, this system also belongs to class D in the Cartan-Altland-Zirnbauer classification scheme for topological superconductors~\cite{Altland1997Nonstandard, Sato2017Topological}.

%
We proceed numerically by diagonalizing the Hamiltonian within the BdG formalism and extract the low-energy levels and their corresponding eigenvectors using the iterative Lanczos method, which is applicable because of the sparse BdG Hamiltonian, enabling us to handle lattices corresponding to $\sim 10^4\times 10^4$ matrices for studying the low-energy physics~\cite{Awoga2022Robust,Awoga2023Estimating}. We find qualitatively similar results as for the continuum superconductors in Sec.~\ref{subsec:Cont}, underlining the model independence of our results. In particular, the Majorana hybridization as a function of the spin orientation of the single magnetic adatom shows a clear shielding effect with concomitant changes in the parity of the ground state, see Fig.~\ref{fig:Fig4}(a).

%
%
In Sec.~\ref{subsubsec:BrydonModel}, we defined the $y$ offset of the magnetic adatom $y_{\rm S}^{}<y_{\rm S, max}^{}$ to obtain Majorana shielding, see Fig.~\ref{fig:Fig3}(b). For the lattice model here, we find $y_{\rm S, max}^{}$ to be a few lattice sites away from the chains.  Figure~\ref{fig:Fig4}(b) is a representative plot of the in-gap levels $E_{\rm M}$ and $E_{\rm S}^{}$ as functions of $y_{\rm S}^{}$ for different $x_{\rm S}^{}$. 
The exchange coupling, $J_{\rm S}^{}$, between the single magnetic adatom and the superconductor, also tunes the Majorana hybridization, as shown in Fig.~\ref{fig:Fig4}(c) for two positions of the single magnetic adatom. This is expected since $J_{\rm S}$ determines $\epsilon_{\rm S}^{}$, corroborating the dependence of the ingap states on the bare YSR energy and its control of the shielding effect shown in Fig.~\ref{fig:Fig2}(c). It is worth mentioning that at small and large  $J_{\rm S}^{}$, the bare YSR energy, $\epsilon_{\rm S}^{}$, exceeds the topological gap and thus the effective MZM-YSR coupling is almost vanishing, i.e., $E_{\rm M} \approx \epsilon_{\rm M}$, in that regime. 
%
\section{Experimental considerations}\label{sec:Expt}
In this section, we discuss possible experimental approaches for manipulating and detecting the described Majorana hybridization. A priori, there are several parameters that can be tuned to control the hybridization: the single adatom's position on the substrate, the induced YSR energy, and the orientation of the adatom spin. The interchain distance $R$ is difficult to alter; thus we consider it fixed in our analysis. One of the caveats of the YSR route for designing unconventional superconducting devices is that the magnetic adatoms can be changed on the time scale of milliseconds to seconds, making it impossible to alter this system parameter within the coherence time of the MZM. One way to bypass such parameter rigidity is to trigger the magnetic adatom spin dynamics in ESR-STM setups, a technique that has gained a lot of attention due to its sub-$\mu$eV energy resolution and its ability to detect and manipulate the spin of single atoms ~\cite{Balatsky2012Electron,Baumann2015Electron,DELGADO2021A,Drost2021Combining}. An applied radio-frequency voltage between an STM tip, harboring a spin ${\bs S}_{\rm tip}$, and the substrate induces a variation in the tip-substrate distance, which in turn creates a modulation of the exchange coupling $J_{\rm ex}$ between ${\bs S}_{\rm tip}$  and the spin ${\bs S}_{\rm S}$ of the single magnetic adatom, which effectively generates an $ac$ magnetic field that can drive the latter. While a full description of the adatom spin dynamics is beyond the scope of this paper, here we provide a short account of the effective model that has been used to describe ESR-STM measurements. A minimal adatom spin Hamiltonian consistent with the experimental findings reads \cite{Beck2023Systematic,note}: 
\begin{align}
    \mathcal{H}_{m}(t)&=-KS_{{\rm S},z}^2+\gamma {\bs B}(t)\cdot{\bs S}_{\rm S}\,,
    \label{hamAdatom}
\end{align}
where $K>0$ is the easy-axis anisotropy,  $\gamma$ is the gyromagnetic factor, and ${\bs B}(t)=(J_{\rm ex}(t)/\gamma)\langle {\bs S}_{\rm tip}\rangle+{\bs B}_{\rm ex}$ is the total magnetic field acting on the adatom, being the sum of the time-dependent exchange bias and the externally applied field. Here, $\langle{\bs S}_{\rm tip}\rangle$ is the average spin of the tip, which is determined by both the external magnetic field ${\bs B}_{\rm ex}$, and local anisotropies \cite{Yang2019Tuning}. The equilibrium orientation of the adatom spin, ${\bs S}_{\rm S}(\theta_{\rm S},\phi_{\rm S})$, can be found by minimizing the energy in Eq.~(\ref{hamAdatom}). Since $J_{\rm 
 ex}\propto \exp(-d/d_{\rm ex})$ \cite{Yang2019Tuning}, with $d_{\rm ex}$ a characteristic length,  changing the tip height $d$ could facilitate the rotation of the spin ${\bs S}_{\rm S}$ by arbitrary angles $\{\theta_{\rm S}, \phi_{\rm S}\}$. We note in passing that the term $\sim S_{{\rm S},z}^2$ is generated by breaking the inversion symmetry of the surface states and, at a more fundamental level, enables the use of the classical approximation for describing the adatom spin \cite{Zitko2018,heinrich2018single}.   
 
The same Hamiltonian in Eq.~(\ref{hamAdatom}) can be harnessed for the readout of the electronic state in ESR-STM measurements. To demonstrate this, we first stress that the electronic spin density at the position of the adatom ${\bs R}_S$ can act back on the spin ${\bs S}_{\rm S}$ via its exchange coupling $J$, or the effective magnetic field ${\bs B}_{n}=(J/\gamma)\langle{\bs \sigma}({\bs R}_{\rm S})\rangle_n$, where $\langle{\bs \sigma}({\bs R}_{\rm S})\rangle_n\equiv\langle\psi_n|{\bs \sigma}({\bs R}_{\rm S})|\psi_n\rangle$ represents the spin expectation value of the superconductor condensate at the position of the adatom in the many-body state $|\psi_n\rangle$. Consequently, the dynamics of the adatom spin depends on the electronic state via the total magnetic field
\begin{align}
    {\bs B}_{n,\rm tot}={\bs B}+{\bs B}_n\,,
\end{align}
which in turn influences ${\bs S}_{\rm S}\equiv{\bs S}_{{\rm S},n}$ and hence the resonance condition. The jiggling of the STM tip height $d$ at a frequency $\omega$ would give rise to a time-dependent magnetic field $\bs B_{\rm ac}(t)\perp{\bs S}_{{\rm S},n}\propto\delta J_{\rm ex}\cos(\omega t)$, with $\delta J_{\rm ex}$ the amplitude of the $ac$ field. Then, sweeping $\omega$ should allow extracting the electronic state-dependent resonance frequency in ESR-STM measurements \cite{Baumann2015Electron,mishra2021dynamical,Mishra2021Yu}. 

Another viable way to accomplish full control of the hybridization is to
place a single magnetic adatom directly on a superconducting STM tip, instead of the bulk superconductor. Its interaction with the bulk superconductor (and hence its YSR energy) can be tuned by changing the height of the STM tip, while its placement on top of the superconductor can be arbitrary ~\cite{Farinacci2018Tuning,Drost2021Combining,Karan2022Superconducting,Schulte2023Changing}. The ESR-STM technique would work analogously to the intrinsic adatom case described above, although the microscopic coupling parameters and their scaling with the STM position would need separate investigation. 

The detection of the ESR signal is performed in conductance measurements, and thus the effective $ac$ fields that drive the impurity pertain to charge currents flowing into the YSR state, causing relaxation and dephasing via inelastic electron tunneling. Hence, both manipulation and detection must be performed on time scales shorter than the induced decoherence time $T_2$. Time scales on the order of $T_2\sim 10^2\mu$s have recently been reported for adatoms on normal metals~\cite{Baumann2015Electron}, which also makes this technique particularly attractive also for superconducting substrates. 

So far, we have addressed the static and dynamical aspects of the proposed setup in Fig.~\ref{fig:Fig1}. Next, we consider the non-Abelian aspects, i.e., braiding and fusion manipulations. Paramount to these experiments are the time constraints of these operations, which we discuss below.
%
\subsubsection*{Fusion and braiding operation conditions}\label{subsubsec:OpCond}
For braiding and fusion of Majorana modes, the two most important constraints on the operation time $T$ are the following \cite{aasen2016milestones,vanHeck2012Coulomb,Amorin2015Majorana}: the operation should be sufficiently slow to avoid bulk excitations which gives a lower limit, $T_{\rm adiabatic}$ and it should be fast enough such that the Majorana modes are degenerate within this time scale setting an upper limit, $T_{\rm splitting}$. Braiding protocols require at least three chains connected in a trijunction. However, two chains, as discussed in the above sections, are sufficient for a fusion operation~\cite{aasen2016milestones}. During braiding, for a trijunction protocol \cite{vanHeck2012Coulomb,tsintzis2023roadmap,luna2023design}, two shielded Majoranas are uncoupled or very weakly coupled with hybridization energy, $E_{\rm M}^{\rm min}$, while the remaining Majoranas are strongly coupled with hybridization energy, $E_{\rm M}^{\rm max}$. Thus, in general, the relevant energy that determines the slowest time scale, $T_{\rm splitting}$, is $E_{\rm M}^{\rm min}$, while the fastest time scale, $T_{\rm adiabatic}$, takes into consideration the energy levels of strongly coupled Majorana modes. It is worth mentioning that in fusion protocols that require two chains, there are no extra unshielded pairs of MZMs that need to be taken into account~\cite{seoane2022fusion}.

\begin{figure*}[!t]
	\includegraphics[width=1\textwidth]{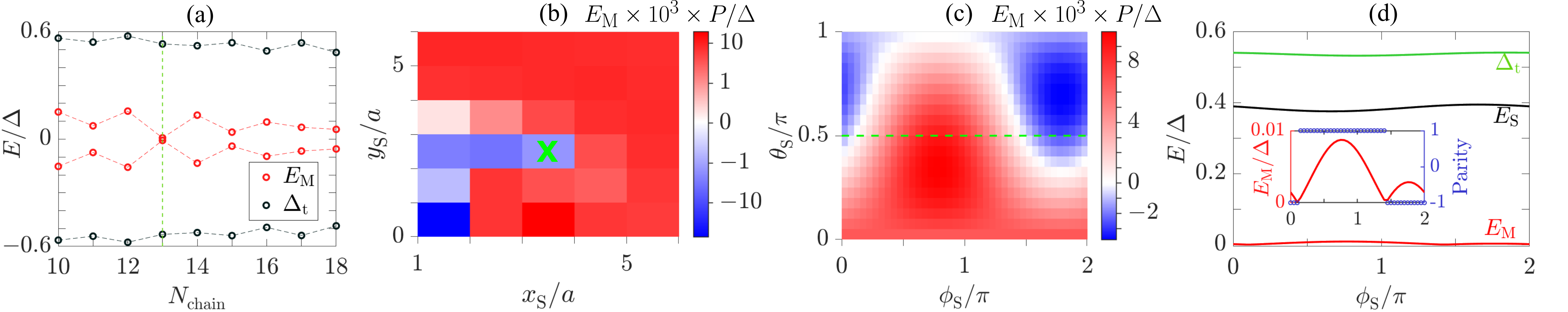}
	\caption{Application of our shielding mechanism to precursors of Majorana modes using a short single ferromagnetic chain on a 2D square lattice $s$-wave superconductor with finite spin-orbit coupling.
		(a) Majorana hybridization in the absence of the single magnetic adatom, $E_{\rm M}^{}=\epsilon_{\rm M}^{}$,  as a function of chain length $N_{\rm chain}^{}$. The dashed green line notes a chain length for which the hybridization is almost zero.
		(b)  Majorana hybridization $E_{\rm M}^{}\times P$, for $N_{\rm chain}^{}=13$ (dashed green line in (a)), as a function of the single impurity position at the right end of the chain for $\left(\theta_{\rm S}^{},\phi_{\rm S}^{}\right)=\left(\pi/2,0\right)$.  Note the logarithmic scale of the colorbar.
		(c) $E_{\rm M}\times P/\Delta$ as a function of the orientation of a single magnetic impurity placed at $\left(x_{\rm S}^{},y_{\rm S}^{}\right)=\left(3a,2a\right)$, marked ``green X" in (b). Compare to Figs.~\ref{fig:Fig2}(a),~\ref{fig:Fig3}(a), and~\ref{fig:Fig4}(a).
		(d) $E/\Delta$ as a function of $\phi_{\rm S}^{}$ for $\theta_{\rm S}^{}=\pi/2$ (along the dashed green line in (c)). Only positive $E_{\rm M}$ (red), $E_{\rm S}$ (black) and $\Delta_{\rm t}$ (green) are shown for clarity. Inset: Close-up of $E_{\rm M}$ (left axis), and the corresponding ground-state parity (right axis), clearly showing the behavior of the hybridization and the parity flip at zero crossing. 
		The calculation is carried out on an $39\times 29$ square lattice with $\Delta=0.5t,\, \lambda =0.4t,\, \mu = 0.5t,\, t=1$, for fixed $J/\Delta= 6.28$, $J_{\rm S}^{}/\Delta= 4.4$.}
	\label{fig:Fig5}
\end{figure*}
In our setup, the energy of the level closest to the Majorana can be one of the following: the topological gap, $\Delta_{\rm t}^{}$, the hybridized YSR energy, $E_{\rm S}^{}$ and $E_{\rm M}^{\rm max}$ (for braiding protocols). Therefore, the operation window is:
\begin{equation}\label{eq:OpTime}
    \dfrac{\hbar}{\min \lbrace \Delta_{\rm t},E_S,E_{\rm M}^{\rm max}\rbrace - E_{\rm M}^{\rm min}}  \ll T \ll \dfrac{\hbar}{E_{\rm M}^{\rm min}}.
\end{equation}We have shown, in Fig.~\ref{fig:Fig2}(a), \ref{fig:Fig3}(a) and \ref{fig:Fig4}(a), that $E_{\rm M}^{\rm min}$ can be tuned very close and also exactly to zero and thus, the experimental accuracy sets the slowest time limit. Therefore, we consider $ \min\lbrace\Delta_{\rm t},E_S,E_{\rm M}^{\rm max}\rbrace-E_{\rm M}^{\rm min} \approx\min\lbrace\Delta_{\rm t},E_S,E_{\rm M}^{\rm max}\rbrace.$
In magnetic adatom-superconductor systems, the topological gaps can potentially be large, $\Delta_{\rm t}\sim \left[ 10^{-2} \Delta,10^{-1} \Delta\right]$, for instance in Mn chains on Nb(110)~\cite{schneider2021topological}.  Also, $E_S$ can be tuned to values even larger than $\Delta_{\rm t}$, [see Fig.~\ref{fig:Fig3}(c)] and  $E_{\rm M}^{\rm max}$ can go up to $ 10^{-2}\Delta$ [see Fig.~\ref{fig:Fig4}(a)]. Thus, we consider ${\min \lbrace \Delta_{\rm t},E_S,E_{\rm M}^{\rm max}\rbrace} \sim 10^{-2}\Delta \gg E_{\rm M}^{\rm min}$. This gives the lower limit for operation time to be $10^{-11}{\rm~ s}$, that is, the upper limit of frequency is $100{\rm GHz}$. In practice, a large ratio $E^{\rm max}_{\rm M}/E^{\rm min}_{\rm M}$ allows for a low accuracy threshold for manipulating the control parameters and enhances the protocol's robustness, see  Fig~\ref{fig:MirroSymThetaFrac}. Operating within the above-estimated frequency constraints is well within the capabilities of ESR-STM, which typically reaches frequencies up to 40 GHz~\cite{Drost2021Combining}. 

\section{Concluding remarks}\label{sec:Conc}
In this work, we have demonstrated efficient control of the hybridization of Majorana zero modes in magnetic chains-superconductor systems by using an additional magnetic adatom in the proximity of the chain. We found that the adatom parameters, such as the spin orientation, the exchange coupling to the superconductor, and its position on the substrate, can alter the Majorana spectrum. Specifically, we have shown that the Majorana shielding, i.e., the complete suppression of the Majorana hybridization, and ground state parity flipping are not only possible but are universal and model independent.  Experimentally, manipulation of the single magnetic adatom parameters and hence the MZMs hybridization could be implemented by using STM and ESR-STM techniques.
%
%
Therefore, our study opens a potential route toward experiments that directly probe the non-Abelian character of MZMs in magnetic chain-superconductor platforms. Moreover, it could facilitate the implementation of a universal set of quantum gates with Majorana qubits. Furthermore, successful control of Majorana hybridization is an essential prerequisite for fusion and several braiding schemes of MZMs~\cite{Burrello2013Braiding,Burnell2014Correlated,seoane2022fusion}. To this end, we envision networks of chain-adatom-chain constructions that allow for the time-dependent tuning of multiple MZM hybridizations at junctions with additional single magnetic adatoms.

In the remainder of the discussion, we emphasize two potential applications of our results beyond the aforementioned ones. Recently, STM experiments have constructed magnetic adatom chains on a superconductor atom by atom and have probed the chains at both ends.
However, the chains achieved so far are not long enough, leading to strongly hybridized MZMs, termed precursors of Majorana modes, whose hybridization energy oscillates with chain length~\cite{Schneider2022}. Our proposal shows a path to alleviate this problem. By relaxing the condition that the MZMs on the same chain do not couple and focusing on a single chain, our proposal reduces to the short chain-superconductor system with precursors of Majorana modes, which can be tuned to zero energy by placing a single magnetic adatom close to one end of the chain,
following the procedures of our proposal, see Fig.~\ref{fig:Fig5}. The Majorana hybridization in the absence of the magnetic adatom oscillates with the chain length similarly to what is observed in experiment [see Fig.~4(g-h) in Ref.~\cite{Schneider2022}], see Fig.~\ref{fig:Fig5}(a). By placing the single magnetic adatom near one end of the chain and varying its position or orientation, the ground state parity can be flipped indicating that the hybridization crosses zero, see Fig.~\ref{fig:Fig5}(b-d). 
 Remarkably, a large gap separating the precursors Majorana modes and the next energy level can be engineered, see Fig.~\ref{fig:Fig5}(d). 
Our proposal suggests the use of the shielding effect as a building block for fusion and braiding experiments in magnetic adatom chain systems. By considering realistic experimental parameters, we have provided an estimate for the relevant time scales in fusion and braiding operations.
Thus, our proposal can be tested with state-of-the-art experiments in existing devices. 
The most direct test would be to position an additional magnetic adatom at various positions close to a chain that accommodates precursors of Majorana modes with finite energy and determine the ideal position for Majorana shielding by minimizing the observed energy of the precursors the STM spectrum. 

Another direction for future research concerns the quantum nature of the adatoms spin and its effects on the Majorana hybridization.  Indeed, while here we have used a classical description of the spins which is valid in the limit of large spins and/or large easy axis anisotropies, recent works suggest that quantum effects can become important for smaller spins \cite{OppenPRL22,OppenPRB22,Paaske2023}. In particular, the Kondo-like screening of the adatoms' spins by the Shiba electrons could have consequences on the Majorana hybridization, and on the topological phase diagram in general in our model. Hence, it would be interesting to explore the resulting dynamics pertaining to the coupling between the two quantum systems and eventually establish efficient ways to control the Majorana hybridization in ESR-STM experiments \cite{heinrich2018single}. 
%
\begin{acknowledgments}
OA and ML acknowledge funding from NanoLund, the Swedish Research Council (Grant Agreement No. 2020-03412) and the European Research Council (ERC) under the European Union’s Horizon 2020 research and innovation programme under the Grant Agreement No. 856526. Their simulations were enabled by resources provided by the Swedish National Infrastructure for Computing (SNIC) at the Uppsala Multidisciplinary Center for Advanced Computational Science (UPPMAX), Rackham cluster, partially funded by the Swedish Research Council. I.I acknowledges support
by the Cluster of Excellence “CUI: Advanced Imaging of
Matter” of the Deutsche Forschungsgemeinschaft (DFG)
– EXC 2056 – project ID 390715994 and  T. P. acknowledges funding by the DFG
(project no. 420120155) and the European Union (ERC,
QUANTWIST, project number 101039098). Views and
opinions expressed are however those of the authors only
and do not necessarily reflect those of the European Union
or the European Research Council. Neither the European
Union nor the granting authority can be held responsible
for them. AM and MT thank the Foundation for Polish Science through the International Research Agendas (IRA) program cofinanced by the European Union within Smart Growth Programme. They acknowledge the access to the computing facilities of the Interdisciplinary Center of Modeling at the University of Warsaw, Grant G84-0. AM thanks the University of Delhi for providing financial assistance through the Faculty Research Programme Grant IoE (ref no./2023-24/12/FRP).

\end{acknowledgments}
  
\onecolumngrid 
\appendix
%
\section{Local density of states of the effective low-energy model model}\label{app:LDOS}
In Sec.~\ref{sec:Sys} of the main text, we mention the dependence of the localization of the low-energy states on the parameters of the single magnetic adatom. Here, we present a general analytical derivation of the Majorana weight, $W_{\rm M}^{}$, which is the local density of states of the MZMs at the position of the single magnetic adatom, $d$. We obtain the Green's function of the effective low-energy Hamiltonian, Eq.~\eqref{eq:EffectiveH} in the main text, by solving the Green's functions equations of motion in the basis $\left(   
f,d,f^\dagger,d^\dagger \right)$  given above Eq.~\eqref{eq:EffectiveE} in the main text. The electronic component of the single magnetic adatom's Green's function is
\begin{equation} \label{Greens Function}
    G_{d d^{\dagger}}(E)=\dfrac{A(E)}{(E^2-E_{S}^2)(E^2-E_{M}^2)},
\end{equation}
where we have defined
\begin{equation}
    A(E)=(E^2-\epsilon_{\rm M}^2)(E+\epsilon_{\rm S})-(E-\epsilon_{\rm M})|t_1|^2 -(E+\epsilon_{\rm M})|t_2|^2,
\end{equation}
where $t_1=t_L-it_R$ and $t_2=-\left(t^*_L-it^*_R\right)$. As a consistency check, we observe that the poles of the Green's function appear at the effective energies of Eq. (\ref{eq:EffectiveE}). This yields the local density of states as
%
   {\rm LDOS}($E$)=
    $- \lim_{n \to 0} \Im{G_{d d^{\dagger}}(E+i n)}/\pi.$
%
We find 
\begin{equation}
   {\rm LDOS}(E)=\dfrac{ A(E_M)\delta(E-E_M)+ A(-E_M)\delta(E+E_M)-A(E_S)\delta(E-E_S)- A(-E_S)\delta(E+E_S)}{2 E (E_M^2-E_S^2)},
\end{equation}
which, integrated over energy, equals one, as it should for a single particle state. Hence, the Majorana weight at the single magnetic adatom becomes
\begin{equation}
\label{eq:WM1}
    W_{\rm M}= \dfrac{A(E_M)-A(-E_M)}{2(E_M^2-E_S^2)E_M}= \dfrac{E_M^2-\epsilon_{\rm M}^2-|t_1|^2-|t_2|^2}{E_M^2-E_S^2}.
\end{equation}
We can rewrite  Eq.~\eqref{eq:WM1} as 
\begin{equation} \label{weight result}
    W_{\rm M}=\dfrac{1}{2}-\dfrac{\epsilon_S^2-\epsilon_M^2}{2\sqrt{\left((\epsilon_S-\epsilon_M)^2+4\mid t_{\rm L}-i t_{\rm R} \mid^2\right)\left((\epsilon_S+\epsilon_M)^2+4\mid t_{\rm L}+it_{\rm R}\mid^2\right)}}.
\end{equation}
From the above, we check three different limits. First, if $\epsilon_{\rm S}=\epsilon_{\rm M}$ and $| t_{\rm L} |\neq 0$ or $| t_{\rm R} |\neq 0$, the Majorana LDOS at the position of the single YSR adatom is exactly $\dfrac{1}{2}$. Next, in the large hybridization limit, i.e, $| t_1 | ,| t_2 |\gg \epsilon_{\rm S}, \epsilon_{\rm M}$ the Majorana LDOS is again $\dfrac{1}{2}$. On the other hand, in the small hybridization limit, i.e, $\epsilon_{\rm S}, \epsilon_{\rm M} \gg| t_1 | ,| t_2 |$ the Majorana LDOS at the single adatom vanishes. In order to determine the spatial distribution of the MZM at the parameters of Majorana shielding, we calculate the Majorana LDOS at $E_{\rm M}=0$
\begin{equation}\label{weight at shielding}
W_{\rm M,0}=\dfrac{1}{2}-\dfrac{1}{2}\dfrac{\epsilon_{\rm S}^2-\epsilon_{\rm M}^2}{\left( \epsilon_{\rm S}-\epsilon_{\rm M}\right)^2+4\mid t_{\rm 1,0} \mid^2}=\dfrac{1}{2}-\dfrac{1}{2}\dfrac{\epsilon_{\rm S}^2-\epsilon_{\rm M}^2}{\epsilon_{\rm S}^2+\epsilon_{\rm M}^2+4\mid t_{\rm L,0} \mid^2+4\mid t_{\rm R,0} \mid^2}=\dfrac{1}{2}-\dfrac{1}{2}\dfrac{\epsilon_{\rm S}^2-\epsilon_{\rm M}^2}{E_{\rm S,0}^2},
\end{equation}
where $E_{\rm S,0}$ is the YSR effective energy and $t_{1,0}$ the hopping parameter at the Majorana shielding points. The parameters $t_{L,0}$ and $t_{R,0}$ generally depend on $\epsilon_{\rm S}$ and  $\epsilon_{\rm M}$ themselves. On the other hand, $W_{\rm M,0}$ cannot be expressed solely in terms of $\epsilon_{\rm S}$ and  $\epsilon_{\rm M}$. If $|\epsilon_{\rm S}|\gg |\epsilon_{\rm M}|$, the Majorana shielding is complete $W_{\rm M,0}\approx0$. On the other hand, when $1 \gg \epsilon_{\rm S}- \epsilon_{\rm M}>0$, the Majorana LDOS at the position of the single magnetic adatom is maximal and the states are completely mixed $W_{\rm M,0}\approx\dfrac{1}{2}$.
%
\section{Parity change at zero energy crossing \label{app:Parity}}
We mentioned in the main text that the fermion parity of the ground state changes only through a zero-energy level crossing in symmetry class $D$. Due to the small size of the Hamiltonian in Eq.~(\ref{eq:EffectiveH}), we can demonstrate this behavior explicitly and give details on the polarization of the states. 
To this end, we investigate the changes in the ground state parity as the parameters of the single magnetic adatom are tuned. We consider the effective low-energy Hamiltonian, Eq.~\eqref{eq:EffectiveH}, in Fock space in the basis 
$|n_f,n_d\rangle = \left(f^{\dagger}\right)^{n_f}\left(d^{\dagger}\right)^{n_d}|00\rangle$,
where $n_{f/d}=0,1$ is the fermion occupation of the non-local fermion/YSR state, and the parity of the states is defined as 
$P=\left(-1\right)^{n_{f}+n_{d}}$.
The many-body energies in the odd/even subspaces $E_\pm^{\rm o/e}$ are obtained as,
\begin{subequations}\label{eq:Parity}
    \begin{align}
    E_\pm^{\rm e } & = \frac{1}{2} \left[\epsilon_+ \pm \sqrt{\epsilon_+^2 + 4 |t_-|^2}\right], 
    \\
    E_\pm^{\rm o } & = \frac{1}{2} \left[\epsilon_+ \pm \sqrt{\epsilon_-^2 + 4 |t_+|^2}\right],
    \end{align}
\end{subequations}
 and their corresponding eigenstates are $|V_\pm^{\rm e}\rangle= \pm v_{00}|00\rangle+ v_{11}|11\rangle$ and 
 $|V_\pm^{\rm o}\rangle= \pm v_{10}|10\rangle+ v_{01}|01\rangle$
are linear combinations of $|n_f,n_d\rangle$, with amplitudes $v_{n_f n_d}$, within a single fermion parity sector. 

We see that 
$E_-^{\rm o}$ and $E_-^{\rm e}$
are the lower energies and either of them can be the ground state energy, while 
$E_+^{\rm o}$ and $E_+^{\rm e}$
are higher in energy.
Even and odd states cross at the degenerate point
$E_-^{\rm o} = E_-^{\rm e}$ $\left({\rm }~E_+^{\rm o} = E_+^{\rm e}\right)$
 accompanied by a fermion parity switching. From Eq.~\eqref{eq:Parity}, we see that the condition for the many-body ground state degeneracy is the same as the Majorana shielding condition of Eq.~\eqref{eq:EMeq0}.
%
%
 %
 \section{Derivation of generalized low-energy model coupling, $t_{\rm L/R}$}\label{app:tGen}
In Eq.~\eqref{eq:tGen} in Sec.~\ref{subsec:Cont} of the main text, we presented the generalized expression of the coupling between YSR state and the MZMs in the left and right chains, $t_{\rm L/R}$. Here, we give a detailed derivation of these quantities, independently of the microscopic model. We first consider the Hamiltonian that describes the coupling of the left chain to the single magnetic adatom, obtained from the last line of Eq.~(\ref{eq:GenContH}) in the main text, as
\begin{equation}
   \label{H LM q}
			H_{\rm LS}=d^{\dagger}  \sum_{m} h_{N_L+\frac{R_S}{a},m}b_{Lm}+\Delta_{N_L +\frac{R_S}{a},m}b_{Lm}^{\dagger}+d \sum_{m} -h_{N_L+\frac{R_S}{a}, m }b^{\dagger}_{Lm}+\Delta_{N_L +\frac{R_S}{a},m}b_{Lm}.
\end{equation}
We make a basis transformation of Eq. (\ref{H LM q})
\begin{equation} \label{modes}
	H_{\rm LS}=d^{\dagger} \left(t_{\rm L} \gamma_L+ \sum_m q_m p_m\right)+ \left(t^*_L \gamma_L+ \sum_m q^*_m p_m^{\dagger}\right) d,
\end{equation}
where $\gamma_{\rm L},p_i$ correspond to the Majorana operator and the YSR band orbitals of the left chain, respectively, and $q_i$ are unknown matrix elements. Next, we extract the matrix element $t_{\rm L}$ via
\begin{equation} \label{t2}
	t_{\rm L}=\dfrac{1}{2} \{d,[H_{\rm LS},\gamma_{\rm L}]\}.
\end{equation}
 By expressing the Majorana solution in the fermionic basis in the second quantized formalism, Eq.~\eqref{t2} yields
  \begin{equation}\label{AppendixEqtL}
      t_{\rm L}=\sum_{m}\left[ u_m h_{N_L +\frac{R_S}{a}, m} + v_m \Delta_{N_L +\frac{R_S}{a}, m}\right],
  \end{equation}
 where $u_m$ and $v_m$ are the particle and hole components of the Majorana solutions' of the Bogoliubov-de Gennes (BdG) eigenmode. We  evaluate the matrix elements $h_{N_L +\frac{R_S}{a}, m}$ and $\Delta_{N_L +\frac{R_S}{a}, m}$ of Eq.~\eqref{eq:GenContMatrixElement} from the following equations,
 \begin{equation}
     \begin{split}
       \langle \uparrow(N_L +\frac{R_S}{a} ) \mid \uparrow\left(m\right)\rangle= &\cos(\theta_{\rm S}^{}/2) e^{i\phi_{\rm S}^{}/2}\cos(\theta_m/2)e^{-i \phi_{\rm m}^{}/2}+\sin(\theta_{\rm S}^{}/2) e^{-i\phi_{\rm S}^{}/2}\sin(\theta_m/2)e^{i \phi_{ m}^{}/2}, \\
    \langle \uparrow(N_L +\frac{R_S}{a} ) \mid \downarrow \left(m\right) \rangle=&\cos(\theta_{\rm S}^{}/2) e^{i\phi_{\rm S}^{}/2}\sin(\theta_m/2)e^{-i \phi_{ m}^{}/2}-\sin(\theta_{\rm S}^{}/2) e^{-i\phi_{\rm S}^{}/2}\cos(\theta_m/2)e^{i \phi_{m}^{}/2},\\
     \langle \uparrow(N_L +\frac{R_S}{a} ) \mid i\sigma_y \mid \uparrow\left(m\right)\rangle= &\cos(\theta_{\rm S}^{}/2) e^{i\phi_{\rm S}^{}/2}\sin(\theta_m/2)e^{i \phi_{m}^{}/2}-\sin(\theta_{\rm S}^{}/2) e^{-i\phi_{\rm S}^{}/2}\cos(\theta_m/2)e^{-i \phi_{m}^{}/2},\\
     \langle \uparrow(N_L +\frac{R_S}{a} )\mid i\sigma_y \mid \downarrow\left(m\right) \rangle= &-\cos(\theta_{\rm S}^{}/2) e^{i\phi_{\rm S}^{}/2}\cos(\theta_m/2)e^{i \phi_{m}^{}/2}-\sin(\theta_{\rm S}^{}/2) e^{-i\phi_{\rm S}^{}/2}\sin(\theta_m/2)e^{-i \phi_{m}^{}/2}.
     \end{split}
 \end{equation}
By substituting the matrix elements in Eq. (\ref{AppendixEqtL}), the general form of $t_{\rm L}$ is expressed as 
 \begin{equation}\label{resulttL}
     t_{\rm L} =e^{i\frac{\phi_{\rm S}^{}}{2}}\cos\left(\frac{\theta_{\rm S}^{}}{2}\right)\, F_{L} + e^{-i\frac{\phi_{\rm S}^{}}{2}}\sin\left(\frac{\theta_{\rm S}^{}}{2}\right)\, G_{\rm L},
 \end{equation}
 which is presented in Eq.~\eqref{eq:tGen} in Sec.~\ref{subsec:Cont} of the main text. Here,
  \begin{equation}\label{eq:AppFG}
     \begin{split}
    F_{\rm L}=&\dfrac{1}{2}\sum_{m=0}^{\infty} u_m \left(h^{(0)}_{N_L +\frac{R_S}{a}, m}\cos(\theta_m/2)e^{-i \phi_{\rm m}^{}/2}+h^{(1)}_{N_L +\frac{R_S}{a},m}\sin(\theta_m/2)e^{i \phi_{\rm m}^{}/2}\right) \\
    & +v_m \left(\Delta^{(0)}_{N_L +\frac{R_S}{a}, m} \sin(\theta_m/2)e^{-i \phi_{\rm m}^{}/2}-\Delta^{(1)}_{N_L +\frac{R_S}{a}, m}cos(\theta_m/2)e^{i \phi_{\rm m}^{}/2}\right),\\
    G_{\rm L}=&\dfrac{1}{2}\sum_{m=0}^{\infty} u_m \left(h^{(0)}_{N_L +\frac{R_S}{a},m}\sin(\theta_m/2)e^{i \phi_{\rm m}^{}/2}-h^{(1)}_{N_L +\frac{R_S}{a},m}\cos(\theta_m/2)e^{-i \phi_{\rm m}^{}/2}\right) \\
    & -v_m \left(\Delta^{(0)}_{N_L +\frac{R_S}{a}, m} \cos(\theta_m/2)e^{i \phi_{\rm m}^{}/2}+\Delta^{(1)}_{N_L +\frac{R_S}{a}, m}\sin(\theta_m/2)e^{-i \phi_{\rm m}^{}/2}\right).
     \end{split}
 \end{equation}
Hence, we explicitly show the dependence of the coupling of the YSR state to the MZM on the orientation of the single magnetic adatom. Notably, from Eq.~\eqref{eq:AppFG}, we establish that
\begin{equation}
    F_{\rm L}(\phi_{m}^{},\theta_m)= G_{\rm L}(-\phi_{m}^{},\pi-\theta_m).
\end{equation}
A similar form can be derived for $t_{\rm R}^{}$ by substituting the Majorana solution and the matrix elements of the right chain.

So far, we have derived the general expressions without the microscopic details of the chains and superconductor. We next focus on special cases which are model dependent.
%
\subsection{Special Case: Helical Chain Model} \label{SpecialCaseAppendix}
Let us consider the special case where the spins of magnetic adatoms in the chains form a planar helix with $\theta_m=\theta=\pi/2$ and a chiral symmetry $\mathcal{C}$ is present in the system.This specific case corresponds to the model presented in Sec.~\ref{subsubsec:PientkaModel}. For $\mathcal{C}=\sigma_y$, the zero energy modes of the chains will be chiral counterparts of the $\mathcal{C}$ operator. Here, $\sigma_y$ is a Pauli matrix which acts on the basis of the particle-hole components of the YSR orbitals. This constraints the form of the BdG eigenspinors of the Majoranas to satisfy $v_m=iu_m$ for the $\gamma_{\rm L}$ \cite{Pientka2014Unconventional}. The equality $u^*=v$ holds due to particle-hole symmetry. Under these assumptions, we compute 
 \begin{equation}
    F_{\rm L}^*=\dfrac{1}{2\sqrt{2}} \sum_{m} e^{i \phi_{\rm m}^{}/2} \left( v_m h^{(0)}_{N_L +\frac{R_S}{a}, m}+u_m \Delta^{(0)}_{N_L +\frac{R_S}{a}, m} \right)=\dfrac{i}{2\sqrt{2}} \sum_{m} u_m  e^{i \phi_{\rm m}^{}/2}\left(h^{(0)}_{N_L +\frac{R_S}{a}, m}-i\Delta^{(0)}_{N_L +\frac{R_S}{a}, m}\right)= i G_{\rm L}.
\end{equation}
We conclude that $G_{\rm L}=-i F^*_{\rm L}$ as mentioned in Sec. \ref{subsubsec:PientkaModel} of the main text. It follows that $G^*_{\rm L}=i F_{\rm L}$ and $|G_{\rm L}|=|F_{\rm L}|$. Similarly, we obtain $F^*_{\rm R}=-i G_{\rm R}$. Using the above, we calculate the absolute value of the coupling $|t_{\rm L}|^2$ from Eq. (\ref{resulttL}). We arrive at
\begin{equation}
|t_{\rm L}|^2=|G_{\rm L}|^2+2\sin\left(\dfrac{\theta_{\rm S}}{2}\right)\cos\left(\dfrac{\theta_{\rm S}}{2}\right) \Re\left[ie^{i\phi_{\rm S}}G^2_L\right].
\end{equation}
Here $|t_{\rm L}|^2$ is symmetric for $\theta_S \rightarrow \pi-\theta_S$, as expected. This derivation is general in the sense that we have not assumed a semi-infinite chain and do not restrict our calculations to the Bragg point $k_F=k_h$.
%
%
\subsection{Special case: Ferromagnetic Chain Model}\label{app:GenFM}
\begin{figure*}[!t]
  \begin{centering}
	\includegraphics[width=0.7\textwidth]{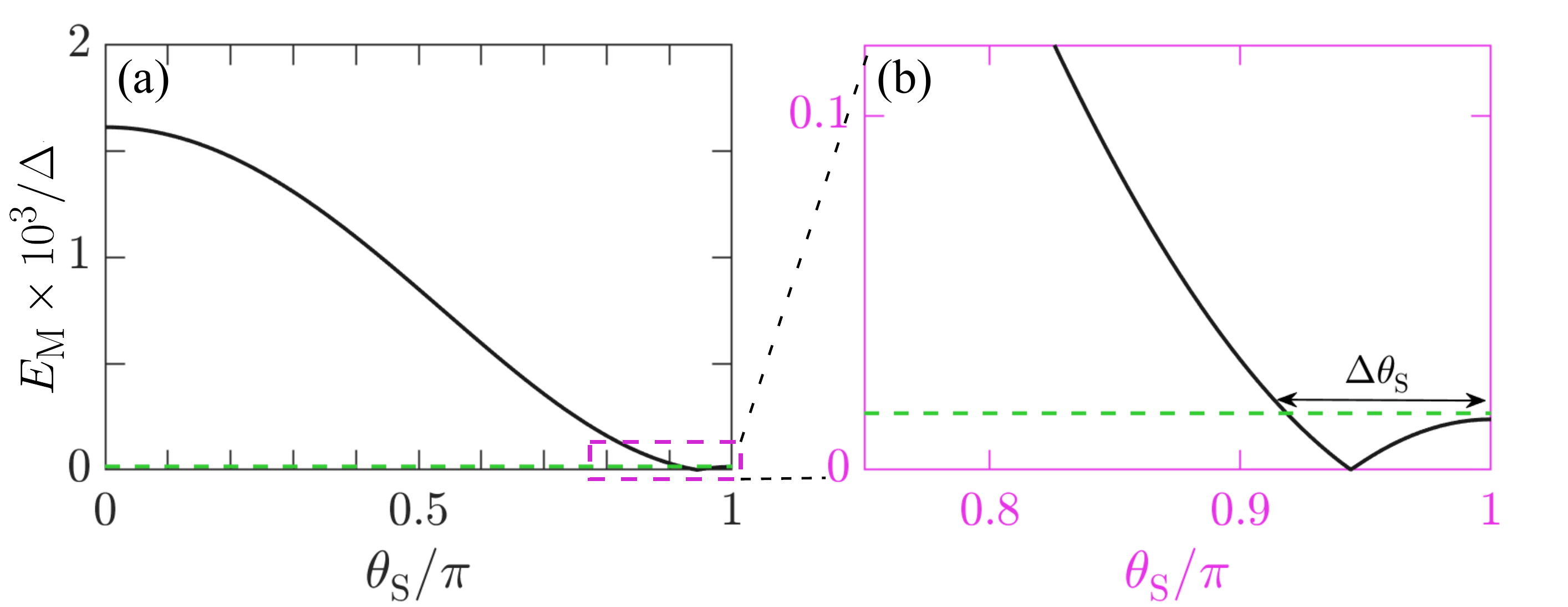}
\caption{Tuning of Majorana energy, $E_M$, from the decoupled regime (below the green line) to the coupled
regime (above the green line) solely by the polar angle of the control atom's magnetization in a mirror-symmetric ferromagnetic impurity chains atomic setup.
(a)  $E_M$ as a function of $\theta_{\rm S}$, and (b) Zoom-in view of the magenta dashed box is given in (a). The single impurity is positioned at the midpoint between the chains ($R_S=10a$), which are separated by a distance $R=20a$. Due to mirror-symmetry, $E_{\rm M}$ is independent of $\phi_{\rm S}$. The green line denotes the condition, $E_M=0.01E_{\rm M}^{\rm max}$. Majorana modes are considered effectively shielded when $E_M$ lies below this threshold, represented by the green line in (a). This gives an error tolerance window $ \dfrac{\Delta\theta_{\rm S}}{\pi}\sim 8\%$ for fine-tuning the control impurity orientation as shown in (b).  Here, $\epsilon_S=0.013\Delta,~\lambda=0.09\Delta, \epsilon_0=0.005\Delta, \xi_{\rm{sc}}=5a,~N_L=N_R=100$.}
	\label{fig:MirroSymThetaFrac}
  \end{centering}
\end{figure*}

We now consider the model presented in Sec.~\ref{subsubsec:BrydonModel}. In the case where the magnetizations of the YSR atoms of the chains are pointing perpendicular to the SOC plane of the substrate in the z direction, the effective BdG Hamiltonians of each of the chains acquire an extra chiral symmetry $\mathcal{\tilde{C}}=\sigma_x$. This implies that the Majorana BdG eigenspinors are restricted to be eigenstates of $\sigma_x$. This implies $u_{\rm L}=-v_{\rm L}$ and $u_{\rm R}=v_{\rm R}$ or $u_{\rm R}=-v_{\rm R}$ and $u_{\rm L}=v_{\rm L}$. Without loss of generality, we focus on the first case. Combining that with the reality condition of the Majorana wavefunction, we conclude that $u_{\rm R}=v_{\rm R}=u$ and $u_{\rm L}=-v_{\rm L}=iu$, where $u$ is real.  We verified the above by a numerical evaluation of the Majorana wavefunctions. In that case, the parameters in Eq.~\eqref{eq:AppFG} can be simplified to
\begin{equation}
 \begin{split}
     F_{\rm L}=&\dfrac{1}{2}\sum_{m=0}^{\infty} u_{Lm} h^{(0)}_{N_L +\frac{R_S}{a}, m}-v_{Lm}\Delta^{(1)}_{N_L +\frac{R_S}{a}, m}=\dfrac{i}{2}\sum_{m=0}^{\infty} u_{m} \left( h^{(0)}_{N_L +\frac{R_S}{a}, m}+\Delta^{(1)}_{N_L +\frac{R_S}{a}, m}\right),\\
    G_{\rm L}=&-\dfrac{1}{2}\sum_{m=0}^{\infty} u_{Lm} h^{(1)}_{N_L +\frac{R_S}{a}, m}+v_{Lm}\Delta^{(0)}_{N_L +\frac{R_S}{a}, m}=-\dfrac{i}{2}\sum_{m=0}^{\infty} u_{m} \left( h^{(1)}_{N_L +\frac{R_S}{a}, m}-\Delta^{(0)}_{N_L +\frac{R_S}{a}, m}\right),
\end{split}
\end{equation}
where the matrix elements$h^{0(1)},\Delta^{0(1)}$ are calculated in Appendix \ref{app:FMeffective}. We write the same expressions for $t_{\rm R}$
\begin{equation}\label{eq:appGFferro}
 \begin{split}
      F_{\rm R}=&\dfrac{1}{2}\sum_{m=0}^{\infty} u_{Rm} \tilde{h}^{(0)}_{N_L +\frac{R_S}{a}, m}-v_{Rm}\tilde{\Delta}^{(1)}_{N_L +\frac{R_S}{a}, m}=\dfrac{1}{2}\sum_{j=0}^{\infty} u_{m} \left( \tilde{h}^{(0)}_{N_L +\frac{R_S}{a}, m}-\tilde{\Delta}^{(1)}_{N_L +\frac{R_S}{a},m}\right),\\
    G_{\rm R}=&-\dfrac{1}{2}\sum_{m=0}^{\infty} u_{Rm} \tilde{h}^{(1)}_{N_L +\frac{R_S}{a}, m}+v_{Rm}\tilde{\Delta}^{(0)}_{N_L +\frac{R_S}{a}, m}=-\dfrac{1}{2}\sum_{m=0}^{\infty} u_{m} \left( \tilde{h}^{(1)}_{N_L +\frac{R_S}{a},m}+\tilde{\Delta}^{(0)}_{N_L +\frac{R_S}{a}, m}\right),
\end{split}
\end{equation}
where the tilded matrix elements couple the single magnetic adatom to the right chain and the index j runs over the right chain. We now focus on the case where the single magnetic adatom is placed exactly in the middle of the two chains $R_{\rm S}=R/2$. Notably, the elements $h^{1(0)}$ and $\Delta^{1(0)}$ are odd( even) functions of the relative positions of the YSR states. We rewrite the matrix elements Eq.~\eqref{eq:appGFferro}
\begin{equation}
 \begin{split}
      F_{\rm R}=&\dfrac{1}{2}\sum_{m=0}^{\infty} u_{m} \left( h^{(0)}_{N_L +\frac{R_S}{a}, m}+\Delta^{(1)}_{N_L +\frac{R_S}{a} ,m}\right),\\
    G_{\rm R}=&=-\dfrac{1}{2}\sum_{m=0}^{\infty} u_{m} \left( -h^{(1)}_{N_L +\frac{R_S}{a} ,m}+\Delta^{(0)}_{N_L +\frac{R_S}{a}, m}\right).
\end{split}
\end{equation}
We establish that $F_{\rm L}-i F_{\rm R}=0$ and also $G_{\rm L}+i G_{\rm R}=0$. From Eq. (\ref{eq:EffectiveE}), it is evident that $\phi_{\rm S}$ enters $E_{\rm M}$ in the terms $t^2_{\pm}=|t_{\rm L}\mp i t_{\rm R}|^2$. Thus, we compute 
\begin{align}
    |t_{\rm L}+i t_{\rm R}|=&|e^{i\phi_{\rm S}}\cos\left(\dfrac{\theta_S}{2}\right)\left(F_{\rm L}+i F_{\rm R} \right)|=|\cos\left(\dfrac{\theta_S}{2}\right)\left(F_{\rm L}+i F_{\rm R} \right)|,\\
     |t_{\rm L}-i t_{\rm R}|=&|e^{-i\phi_{\rm S}}\sin\left(\dfrac{\theta_S}{2}\right)\left(G_{\rm L}-i G_{\rm R} \right)|=|\sin\left(\dfrac{\theta_S}{2}\right)\left(G_{\rm L}-i G_{\rm R} \right)|.
\end{align}
 Both the above expressions are independent of $\phi_{\rm S}$. Thus, we proved that when the single magnetic adatom is placed at the middle of the chains, its' azimuthal orientation does not influence the hybridization of the MZMs. see Fig.~\ref{fig:MirroSymThetaFrac} for a numerical demonstration.
%
  %
 %
\section{Derivation of the effective model parameters for \ref{subsubsec:PientkaModel}} \label{app:ToyModelParams}
In this Section, we present a detailed derivation of the low-energy parameters, $\epsilon_{\rm M},\, t_{\rm L}$ and $t_{\rm R}$, in Sec.~\ref{subsubsec:PientkaModel} of the main text. In the following, the semi-infinite chain limit and the Bragg point approximation  are considered.  
Following \cite{Pientka2014Unconventional}, we rotate in the fermionic basis and express the normalized Majorana operators at the right edge of the left chain $\gamma_{\rm L}$ and at the left edge of the right chain $\gamma_{\rm R}$ as
\begin{align} \label{MajoranaopL}
		\gamma_{\rm L}=&\sqrt{1-\beta^2} \sum_{m=0}^{\infty}e^{-ln(|\beta|)m}\left(e^{i\pi/4}b_{Lm}+e^{-i\pi/4}b^{\dagger}_{Lm}\right),\\
    \label{MajoranaopR}
		\gamma_{\rm R}=&\sqrt{1-\beta^2} \sum_{m=0}^{\infty}e^{-ln(|\beta|)m}\left(e^{-i\pi/4}b_{Rm}+e^{i\pi/4}b^{\dagger}_{Rm}\right),
\end{align}
		written in the basis of the YSR creation and annihilation operators of the left and right chain, respectively, and $\beta$ is defined in the main text in Eq. (\ref{eq:LowEPientkaPar}). We note the usual exponential decay of the Majorana operators.
%
\subsection{YSR-Majorana couplings, $t_{\rm L},\,t_{\rm R}$} \label{appendix t1}
		Following \cite{Pientka2013topological}, we acquire
		\begin{align}
		     \label{h0j}
				h_{N_L +\frac{R_S}{a}, m}=&-\dfrac{\Delta}{\sqrt{2}} \dfrac{\sin(k_F (R_{\rm S}^{}+am))}{k_F (R_{\rm S}^{}+am)} e^{-(R_{\rm S}^{}+am)/\xi_{\rm sc}}\left(\cos\left(\dfrac{\theta_{\rm S}^{}}{2}\right)e^{i(\phi_{\rm S}^{}+2k_ham)/2}+\sin\left(\dfrac{\theta_{\rm S}^{}}{2}\right) e^{-i(\phi_{\rm S}^{}+2k_ham)/2} \right),
				\\
    \label{D0j}
		\Delta_{N_L +\frac{R_S}{a} ,m}=&\dfrac{\Delta}{\sqrt{2}} \dfrac{\cos(k_F (R_{\rm S}^{}+am))}{k_F (R_{\rm S}^{}+am)} e^{-(R_{\rm S}^{}+am)/\xi_{\rm sc}} \left(\cos\left(\dfrac{\theta_{\rm S}^{}}{2}\right)e^{i(\phi_{\rm S}^{}+2k_h a m)/2}-\sin\left(\dfrac{\theta_{\rm S}^{}}{2}\right) e^{-i(\phi_{\rm S}^{}+2k_ham)/2} \right).
		\end{align} 
We now substitute the Majorana solution from Eq. (\ref{MajoranaopL}) in the general Eq. (\ref{resulttL}) to get
\begin{equation}
	t_{\rm L}=\dfrac{e^{-i\pi/4}\sqrt{1-\beta^2}}{2}\sum_m  \beta^m\left(h_{N_L +\frac{R_S}{a}, m}+i \Delta_{N_L +\frac{R_S}{a}, m}\right).
\end{equation}
After substituting the matrix elements  Eq. (\ref{h0j}) and Eq. (\ref{D0j}) and performing this infinite sum, we reach Eq. (\ref{eq:LowEPientkaPar}) of the main text. A similar calculation can be performed for $t_{\rm R}$ where the helicity changes sign $k_h\rightarrow{-k_h}$ in Eq. (\ref{h0j}) and also the Majorana solution $\gamma_{\rm R}$ in Eq. (\ref{MajoranaopR}) should be used.

\subsection{Bare Majorana hybridization, $\epsilon_{M}$} \label{app:BareMajE}
In order to calculate the bare Majorana hybridization between the inner MZMs of the chains, we consider the effective Hamiltonian of Eq.~(\ref{eq:GenContH})
\begin{equation} \label{IntSecQ}
H_{\rm LR}= \sum_{mn} b^{\dagger}_{Lm} \left( h_{mn} b_{Rn}+ \Delta_{mn} b^{\dagger}_{Rn} \right)+\sum_{mn} b_{Lm}\left(\Delta_{mn} b_{Rn}- h_{mn}b^{\dagger}_{Rn} \right),
\end{equation}
where $b_{Lm}$, $b_{Rn}$ are annihilation operators for electrons at the YSR orbitals in the left and right chains, respectively, and 
\begin{align}\label{hij}
	h_{mn}=&-\cos(k_ha(m+n))\dfrac{\Delta\sin(k_F(am+an+R))}{k_F(am+an+R))} e^{-\frac{am+an+R}{\xi_{\rm sc}}},
    \\ \label{Dij}
\Delta_{mn}=&i\sin(-k_ha(m+n))\dfrac{\Delta\cos(k_F(am+an+R))}{k_F(am+an+R))} e^{-\frac{am+an+R}{\xi_{\rm sc}}} ,
\end{align}
are the matrix elements taken from \cite{Pientka2013topological}. We express Eq. (\ref{IntSecQ}) in the fermionic eigenbasis of the left and right chains 
\begin{equation}\label{rotated}
H_{\rm LR}=\dfrac{i}{2}\epsilon_{\rm M} \gamma_{\rm L}\gamma_{\rm R}+ \sum_{mn}k_{m}\gamma_{\rm L}p_{Rm}+l_{n} \gamma_{\rm R} p_{Ln}+M_{mn}p_{Ln}p_{Rm}+H.c,
\end{equation}
where $p_{Ln},p_{Rm}$ correspond to the electrons of the YSR band orbitals of the left and right chain, respectively. These are independent fermionic operators that anti-commute with each other. Also, $k_m$,$l_n$,$M_{mn}$ are material specific matrix elements. These a priori unknown matrix elements do not need to be calculated because we assume that the associated orbitals of the YSR bands are high in energy compared to the bare Majorana hybridization and hence they play no role in the subsequent calculations for the low-energy theory. It is straightforward to extract the element $i\epsilon_{M}$ from Eq. (\ref{rotated}) as
		\begin{equation}
			i\epsilon_{\rm M}=\dfrac{1}{2} \{ \gamma_{\rm R}, [\gamma_{\rm L},H_{\rm LR}] \}.
		\end{equation}
Now, we substitute the Majorana operators from Eq. (\ref{MajoranaopL}) and Eq. (\ref{MajoranaopR}) and $H_{\rm LR}$ from Eq. (\ref{IntSecQ}) to obtain
\begin{equation}\label{intermediate}
i \epsilon_{\rm M}= (1-\beta^2) \sum_{mn} e^{(m+n)ln(\beta)} \left(\Delta_{mn}+ih_{mn} \right).
\end{equation}
Next, by substituting the matrix elements Eq. (\ref{hij}) and Eq. (\ref{Dij}) and perform the double infinite sum, we acquire $\epsilon_{\rm M}^{}(R)$ in Eq. (\ref{eq:LowEPientkaPar}) of the main text. 
%
\section{Asymptotic limits}\label{app:ShieldingAsymp}
We further analyze the model in Sec. \ref{subsubsec:PientkaModel} by taking the asymptotic limit for the parameters in Eq. (\ref{eq:LowEPientkaPar}). In our parameters, we take the asymptotic limit $R/a\rightarrow \infty$ and $R_{\rm S}^{}/a\rightarrow \infty$ to get
\begin{align} \label{Asymptotic Limit 1}
    {}_2F_1(1, \frac{R}{a}; \frac{R}{a}+1 ; e^{-A_s} ) \rightarrow &\dfrac{1}{1-e^{-A_s}},\\
    \label{Asymptotic Limit 2}
    {}_2F_1(2, \frac{R}{a}+1; \frac{R}{a}+2 ; e^{-A_s} ) \rightarrow & \dfrac{1}{\left(1-e^{-A_s}\right)^2}.
\end{align}
Here, a 3D bulk superconductor has been assumed. For practical purposes, the limits in Eq. (\ref{Asymptotic Limit 1}) and Eq. (\ref{Asymptotic Limit 2}) give a good approximation for our considered setup in Fig. \ref{fig:Fig2}, where $R=20a$ and $\epsilon_{M}$ in Eq. (\ref{eq:LowEPientkaPar}) scales as
\begin{equation}
    \epsilon_{\rm M}^{} \sim \dfrac{e^{-R/\xi_{\rm sc}}}{k_F  R} L(k_F  R),
\end{equation}
where $L(k_F R)$ is the oscillating part. In the dilute adatom limit that we are working with, we expect the bare Majorana hybridization to cross zero multiple times as we are increasing the distance between the chains due to the high frequency of the oscillations $k_Fa\gg1$.  This shows that for big distances until $ R\approx \xi_{\rm sc}$, there is a power law decay for the Majorana-Majorana hybridization. When $R\gg\xi_{\rm sc}$, the exponential decay dominates. In both cases, the oscillations part is present. For the parameters $t_{\rm L}$ and $t_{\rm R}$ that appear in Eq. (\ref{eq:LowEPientkaPar}), only the limit of Eq. (\ref{Asymptotic Limit 1}) enters. The asymptotic limit of Eq. (\ref{ZeroPhi}) for $R_{\rm S}^{}\gg a$ is
\begin{equation}\label{Left Crossing Angle Asymptotic}
    \phi_{\rm S,L}(R_{\rm S}^{})=-2 k_F  R_{\rm S}^{}+ 2 \arctan\dfrac{\sin(2 k_F a)}{-e^{\frac{a}{\xi_{\rm sc}}-ln(\beta)}+  \cos(2 k_F a)}.
\end{equation}
As we see in Eq. (\ref{Left Crossing Angle Asymptotic}), the crossing angle $ \phi_{\rm S,L}(R_{\rm S}^{})$ scales linearly with the distance of the single magnetic adatom from the chain.
 %
 %
\section{Quantitative comparison between analytical and numerical solutions in Sec. \ref{subsec:Cont}} \label{app:Error}
\begin{figure}[!t]
\begin{center}
\includegraphics[width=1\textwidth]{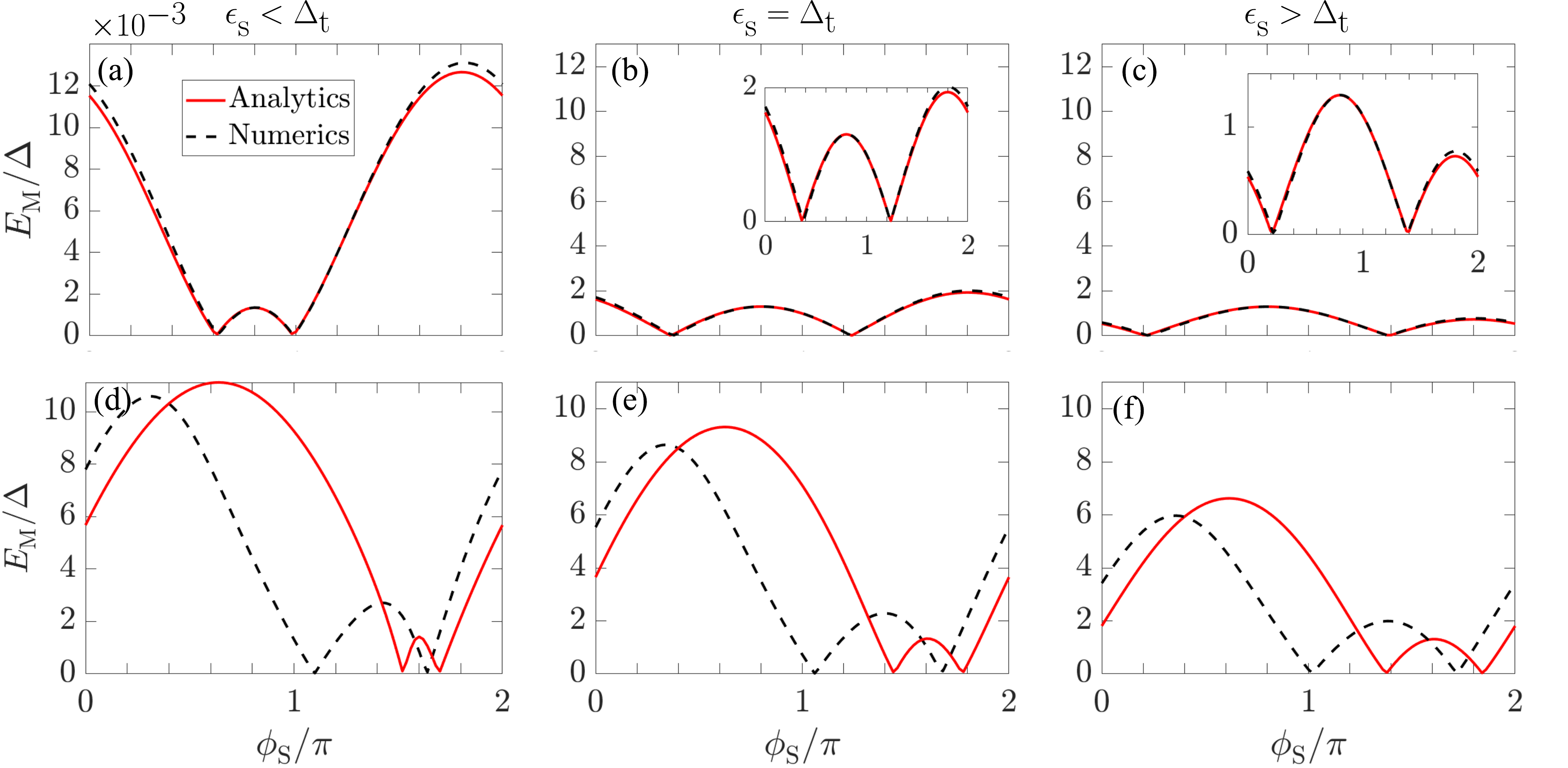} 
\caption{Comparison between the analytical and numerical Majorana hybridization for the helical magnetic spin chain discussed in Section \ref{subsec:Cont} of the main text. (a)-(c) Majorana hybridization, $E_{\rm M}$,  for $\bf{R_{\rm S}^{}}$ $=\left(x_{\rm S}^{},\, y_{\rm S}^{}\right)=\left(8a,0\right)$ within the validity regime of the analytical model. (d)-(f) 
 Same as (a)-(c) but for  $\bf{R_{\rm S}}$ $=\left(x_{\rm S}^{},\, y_{\rm S}^{}\right)=\left(a,0\right)$, outside the validity regime of the analytical model. Here, we consider $\epsilon_{\rm S}^{}=0.01\Delta= 0.2\Delta_{\rm t}^{}$ in (a) and (d), $\epsilon_{\rm S}^{}=0.05\Delta= \Delta_{\rm t}^{}$ in (b) and (e), $\epsilon_{\rm S}^{}=0.08\Delta= 1.6\Delta_{\rm t}^{}$ in (c) and (e). The value of $\Delta_{\rm t}^{}$ is extracted from the numerical solution. We consider $\xi_{\rm SC}^{}=15a$, $R=20a$, $\epsilon_0=0.005 \Delta$, as in Fig. 2 of the main text.}
\label{fig:SpecError} 
\end{center}
\end{figure}
Here, we test the validity of the low-energy approximations and compare the analytical results shown in Fig. \ref{fig:Fig2} with the corresponding numerical solution. The validity of the low-energy model described by Eq. (\ref{eq:EffectiveH}) holds as long as there is small interference between modes of the Shiba bands and the orbital of the single magnetic adatom. In other words, when the couplings $t_{\rm L}^{}$ or $t_{\rm R}^{}$ are of the order of the topological gap, our analytical approximation is no longer valid because second-order effects, due to the interference of the single magnetic adatom with the Shiba bands, can quantitatively change the effective Majorana hybridization $E_{\rm M}^{}$ and YSR energy $E_{\rm S}^{}$. Our derivation of the coupling parameters $t_{\rm L}^{}$ and $t_{\rm R}^{}$ in Eq.~\eqref{modes} has been done under the above approximations.
To test our hypothesis, we compare the analytical and numerical results for two different positions of the magnetic adatom in Fig.~\ref{fig:SpecError}.  When the single magnetic adatom is not too close to one of the chains, $R_{\rm S}$ $\gg a$, exemplified by $\left(x_{\rm S}^{},\, y_{\rm S}^{}\right)=\left(8a,0\right)$ in Fig.~\ref{fig:SpecError}(a)-(c),  numerical and analytical results match also quantitatively, irrespective of the value of $\epsilon_{\rm S}^{}$ as seen in (a), (b), and (c) for $\epsilon_{\rm S}^{} < \Delta_{\rm t}$, $\epsilon_{\rm S}^{} = \Delta_{\rm t}$ and $\epsilon_{\rm S}^{} > \Delta_{\rm t}$, respectively. When the magnetic adatom is close to one of the chains, exemplified by $\left(x_{\rm S}^{},\, y_{\rm S}^{}\right)=\left(a,0\right)$ in Fig.~\ref{fig:SpecError}(d)-(f), we observe strong quantitative deviations between numerical and analytical solutions for $E_{\rm M}^{}$ because the crossing points (as a function of $\phi_{\rm S}^{}$) are shifted. We confirm that on a quantitative level, our analytical formulas are reliable as long as the parameters $t_{\rm L}^{}$ and $t_{\rm R}^{}$ are small enough and there is no significant interference with the Shiba bands. Still, the qualitative picture of the shielding effect (number of zero-energy crossings) is not influenced by the analytical approximations even in the regime of strong quantitative deviations.
%
\section{Ferromagnetic chains on continuum superconductor with non-vanishing spin-orbit coupling}\label{app:FMeffective}
In Sec.~\ref{subsubsec:BrydonModel} of the main text, we presented the results for a model of two ferromagnetic chains in a 2D continuum $s$-wave superconductor with finite Rashba spin-orbit coupling separated by a distance $R$ and a magnetic adatom placed in between at $R_{\rm S}^{}$ away from the left chain. In this section, we give details of the extension of the model in Ref.~\cite{brydon2014topological} to our system.
As discussed in Sec.~\ref{subsec:Cont}, in the dilute deep Shiba limit where the YSR states from individual magnetic adatoms are well within the superconducting gap close to the Fermi energy, the low energy effective Hamiltonian is described by Eq.~\eqref{eq:GenContH}. The matrix elements, $\mathcal{H_{\rm L}},~\mathcal{H_{\rm R}}$ and $\mathcal{H_{\rm LR}}$, for this model can be explicitly written  down using Eq.~\eqref{eq:GenContMatrixElement} as 
\begin{equation}
h_{mn}=h^{(0)}_{mn}=\frac{I^+_1(r_{mn})+I^-_1(r_{mn})}{2},~\Delta_{mn}=\Delta^{(1)}_{mn}=i\frac{I^+_4(r_{mn})-I^-_4(r_{mn})}{2}\,,
\end{equation}
since $\langle \uparrow(m)|\downarrow (n)\rangle=\langle \uparrow(m)|i\sigma_y|\uparrow (n)\rangle=0$ when the magnetic adatoms in the chains are all aligned along z-direction. The matrix elements coupling the single adatom to the chains can be written as 
\begin{align}
h_{N_L+\bs R_{\rm S}^{},m}=&h^{(0)}_{N_L+\bs R_{\rm S}^{},m}\cos\frac{\theta_{\rm S}^{}}{2}e^{i\phi_{\rm S}^{}/2}+ h^{(1)}_{N_L+\bs R_{\rm S}^{},m}\sin\frac{\theta_{\rm S}^{}}{2}e^{-i\phi_{\rm S}^{}/2}\,,\nn\\
\Delta_{N_L+\bs R_{\rm S}^{},m}=& \Delta^{(1)}_{N_L+\bs R_{\rm S}^{},m}\cos\frac{\theta_{\rm S}^{}}{2}e^{i\phi_{\rm S}^{}/2}+  \Delta^{(0)}_{N_L+\bs R_{\rm S}^{},m}\sin\frac{\theta_{\rm S}^{}}{2}e^{-i\phi_{\rm S}^{}/2}\,.
\label{appBrydon:matrixelem}
\end{align}
where $h^{(1)}_{mn}=\frac{I^+_2(r_{mn})-I^-_2(r_{mn})}{2}\frac{x_{mn}+iy_{sm}}{r_{mn}}$ and $\Delta^{(0)}_{mn}=\frac{I^+_3(r_{mn})+I^-_3(r_{mn})}{2}\frac{x_{mn}-iy_{sm}}{r_{mn}}$. In the $k_Fa\gg 1$ limit, the overlap integrals can be written in the asymptotic form as
 \begin{align}
I^\nu_1(r_{mn})\approx&-\mathcal{N}'_\nu(r_{mn})\cos\big(k_{F,\nu}r_{mn}-\frac{\pi}{4}\big),~I^\nu_2(r_{mn})\approx-i~{\rm sgn}(x_{mn})\mathcal{N}'_\nu(r_{mn})\cos\big(k_{F,\nu}r_{mn}-\frac{3\pi}{4}\big),\nn\\
 I^\nu_3(r_{mn})\approx&-\mathcal{N}'_\nu(r_{mn})_\nu\sin\big(k_{F,\nu}r_{mn}-\frac{\pi}{4}\big),~
 I^\nu_4(r_{mn})\approx i~{\rm sgn}(x_{mn})\mathcal{N}'_\nu(r_{mn})\sin\big(k_{F,\nu}r_{mn}-\frac{3\pi}{4}\big).
 \label{appB:asymptotic}
\end{align}
where $\mathcal{N}'_\nu(r_{mn})=\mathcal{N}_\nu\sqrt{\frac{2}{\pi k_{F,\nu}r_{mn}}}e^{-r_{mn}/\xi_{\rm sc}}$ with $\mathcal{N}_\nu=\mathcal{N}_0\big(1-\nu\tilde\lambda/\sqrt{1+\tilde\lambda^2}\big)$, $\mathcal{N}_0$ being the density of states of the superconductor in its normal state at the Fermi level. Here,  $k_{F,\nu}=k_F(\sqrt{1+\tilde\lambda^2}-\nu\tilde\lambda),~v_F=\hbar k_F/\sqrt{1+\tilde\lambda^2}$ with the dimensionless spin-orbit coupling strength $\tilde\lambda=m\lambda/\hbar k_F$ and $\nu=\pm$ representing the helicity sector. Furthermore,  $k_F$ and $v_F$ as the Fermi wave vector and velocity without spin-orbit coupling respectively.
\twocolumngrid
\bibliography{RefsMZMHybridization}

\begin{thebibliography}{114}
\bibitem [{not()}]{note}%
  \BibitemOpen
  \bibinfo {note} {Here we assume the spin $S>1/2$, which is the case for Fe or
  Co adatoms.}\BibitemShut {Stop}%
\end{thebibliography}
%
\end{document}